\documentclass[11pt]{article}
\usepackage[utf8]{inputenc}
\usepackage{amssymb,amsmath}
\usepackage{bm} 
\usepackage{booktabs} 
\usepackage{array}
\usepackage{latexsym}
\usepackage{graphicx}
\usepackage{color}
\usepackage{datetime}
\usepackage[nosort]{cite}
\usepackage{verbatim}
\usepackage{enumerate}
\usepackage{chngpage} 
\usepackage{mathrsfs}
\usepackage{euscript}
\usepackage{psfrag}
\usepackage{subcaption}
\usepackage{tikz}

\usepackage{datetime}

\usepackage[nosort]{cite}
\usepackage{chngpage} 
\usepackage{setspace}
\usepackage{tensor}
\usepackage{physics}

\usepackage{mciteplus}

\usepackage[colorlinks=true,      linkcolor=blue,      urlcolor=blue,      
            filecolor=blue,      citecolor=blue,       pdfstartview=FitH,     
						pdfpagemode=UseNone,      bookmarksopen=true]{hyperref}  
\usepackage[all]{hypcap}     

\definecolor{cardinal}{rgb}{0.6,0,0}
\definecolor{darkgreen}{rgb}{0,0.4,0}
\definecolor{golden}{rgb}{0.92, 0.7, 0}
\definecolor{midnight}{rgb}{0, 0, 0.5}
\definecolor{darkblue}{rgb}{0, 0, 0.7}
\definecolor{purple}{rgb}{0.5, 0, 0.5}

\def\coeff#1#2{\relax{\textstyle \frac{#1}{#2}}\displaystyle}

\def\IR{\mathbb{R}}

\def\ZZ{\mathbb{Z}}

\def\flippt{{\alpha}}

\def\nBPS#1{$\frac{1}{#1}$-BPS}

\numberwithin{equation}{section}

\begin{document}

\phantom{AAA}
\vspace{-10mm}

\begin{flushright}
%
%
\end{flushright}

\vspace{1.9cm}

\begin{center}

{\huge {\bf The M2-M5 Mohawk}}\\

{\huge {\bf \vspace*{.25cm}  }}

\vspace{1cm}

{\large{\bf { Iosif Bena$^{1}$,  Soumangsu Chakraborty$^{1}$, Dimitrios Toulikas$^{1}$  \\ \vskip  3pt 
and  Nicholas P. Warner$^{1,3,4}$}}}

\vspace{1cm}

$^1$Institut de Physique Th\'eorique, \\
Universit\'e Paris Saclay, CEA, CNRS,\\
Orme des Merisiers, Gif sur Yvette, 91191 CEDEX, France \\[12pt]

\centerline{$^3$Department of Physics and Astronomy}
\centerline{and $^4$Department of Mathematics,}
\centerline{University of Southern California,} 
\centerline{Los Angeles, CA 90089, USA}

\vspace{10mm} 
{\footnotesize\upshape\ttfamily iosif.bena @ ipht.fr,  soumangsuchakraborty @ gmail.com, \\
dimitrios.toulikas @ ipht.fr,  warner @ usc.edu} \\

\vspace{2.2cm}
 
 \textsc{Abstract}

\end{center}
\noindent We show that the  near-brane back-reaction of M2 branes ending on M5 branes has a rich ``spike structure'' that is determined by partitioning the numbers of M2 branes that are terminating on groups of M5 branes.  The near-brane limit of the metric describing these branes has an AdS$_3$ factor, implying the existence of  a dual CFT. Each partition of the M2 and M5 charges among spikes gives rise to a different ``mohawk'' revealing a new layer of brane fractionation. We conjecture that all these mohawks are dual to ground states of near-brane-intersection CFT's.  We show that the  supergravity solutions describing these mohawks are part of the large families of AdS$_3$ $\times S^3 \times S^3$ solutions described in   \cite{Bachas:2013vza}.  We identify precisely which of these families are relevant to brane intersections and show that the AdS$_3$ invariance emerges from the self-similarity of the spikes.

\begin{adjustwidth}{3mm}{3mm} 

\vspace{-1.2mm}
\noindent

\end{adjustwidth}

\thispagestyle{empty}
\newpage


\tableofcontents

\section{Introduction}
\label{sec:Intro}

The  study of intersecting brane systems has a vast and diverse history in String Theory.  One can gain significant insight by treating the branes perturbatively, or using the brane actions to determine their dynamics, while the fully back-reacted supergravity solutions for intersecting branes can be very challenging.  This problem can be simplified by smearing the brane distribution, but this can wipe out  essential dynamical details, especially if one is trying to describe black-hole microstructure \cite{Bena:2022sge,Bena:2022ldq,Bena:2022rna}.  Ideally one would like supergravity solutions that describe unsmeared brane intersections, but such solutions typically involve metrics and fluxes that are characterized by complicated systems of non-linear equations.

There are two ways in which such non-linear systems can be rendered manageable.  The first is to try to arrange a very high level of symmetry so that the configuration only depends on one or two variables. In this situation, the equations sometimes simplify to a linear system.  However, such a high level of symmetry often involves smearing out structures that one wants to investigate.  The second approach is to consider some form of ``near-brane'' limit in which some of the functional dependence of the solution is controlled by scale invariance, while the remaining equations can be reduced to a linear system. In this paper we will make a detailed exploration of an example of such a near-brane limit.

We focus on stacks of M2 and M5 branes that intersect along a common $\IR^{1,1}$.  The intersection is thus co-dimension $4$ in the M5 branes, and we will impose spherical symmetry in these directions along the stack of M5 branes.  The intersecting M2-M5 system also has co-dimension $4$ in the complete space time, and we will also impose spherical symmetry in these transverse directions.  The solution therefore has an $\IR^{1,1} \times SO(4)  \times SO(4)$ symmetry, and  depends on three variables, $(u,v,z)$, where $u,v$ are, respectively, radial coordinates in the M5 branes and the transverse space, while $z$ is the remaining coordinate along the M2 branes.  This class of brane intersections has been extensively studied, and the general solution is governed by a non-linear Monge-Amp\`ere-like equation \cite{deBoer:1999gea,Lunin:2007mj,Lunin:2008tf,Bena:2023rzm,Bena:2024qed}.  

{As an offshoot of the study of Janus solutions, a near-brane limit of  intersecting M2 and M5  branes was constructed by seeking out solutions with an $SO(2,2) \times SO(4)  \times SO(4)$ symmetry \cite{DHoker:2008lup,DHoker:2008rje,DHoker:2008wvd,DHoker:2009lky, DHoker:2009wlx,Bobev:2013yra,Bachas:2013vza}.  These solutions contain a warped product of AdS$_3$ $\times S^3 \times S^3$, with the remaining two dimensions described by a Riemann surface, coordinatized by $(\xi, \rho)$.  The underlying BPS equations  were also simplified to a linear system.  In \cite{Bena:2023rzm} it was shown how a class of these solutions could be mapped onto a near-brane limit of the M2-M5 intersections described in \cite{Lunin:2007mj,Lunin:2008tf}.  In particular, it was shown  how, in such a limit, the coordinates $(u,v,z)$ can be recast in terms of the scale coordinate, $\mu$, of a Poincar\'e AdS$_3$ and the coordinates $(\xi, \rho)$ of the Riemann surface.  This work also implicitly implies that, by reducing the problem to only two non-trivial variables,  the  Monge-Ampère-like system can be reduced to a  linear set of equations.}

{Despite this mapping, it remained unclear what kind of M2-M5 systems the solutions of \cite{Bachas:2013vza} describe. The complication is that the solutions of \cite{Bachas:2013vza} involve choices of the Riemann surface and choices of poles and residues in a single function, $G$,  that defines the flux sources. These choices are inextricably linked by requiring regularity of the solution.  In this paper we will make a detailed analysis of certain families of solutions constructed in \cite{Bachas:2013vza} and show that they describe, what appears from infinity, to be a {\it single stack} of semi-infinite M2 branes ending on, and deforming, a {\it single stack}  of M5 branes.  More precisely, the brane stacks are all coincident at infinity but,  because  the M2 branes pull on the M5 branes, the back-reaction causes these stacks to resolve into physically separated spikes ( a ``mohawk''), with the distance between each spike being controlled by the number of  M2's and M5's  making up each spike.}

\begin{figure}[h]
    \centering
    \includegraphics[width=.5 \textwidth]{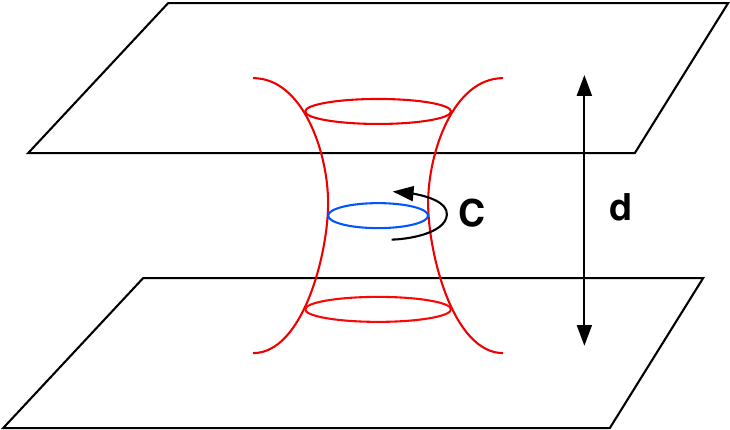}  
    \caption{An M2 brane stretched between two M5 branes.  In the back-reacted solution, the branes will be stretched and the minimum circumference, $C$, around the M2 branes will depend on the separation, $d$, of the M5 branes.}
    \label{Hjunction}
\end{figure}

More generally, we suspect  that any  near-brane limit that leads to such an AdS$_3$ factor is necessarily limited to a single intersection.   The argument  is simple: a more complicated intersection must  involve a scale that would violate the symmetries of the AdS$_3$. Consider, for example, a supergravity solution for an M2 brane stretched between two M5 branes.  As we will see, the supergravity solution for a single intersection faithfully reproduces a geometry consistent with the spike created by an F1 ending on a D-brane \cite{Callan:1997kz, Constable:1999ac}.  An M2 brane stretched between M5's must look like two spikes that meet one another, as depicted in Fig.~\ref{Hjunction},  creating a two-sided  tube between the M5's.  Around this tube there will be a $7$-cycle that is a Gaussian surface for the M2 charge. In the configuration shown in Fig.~\ref{Hjunction}, this $7$-cycle will reach a minimum size at some value of the  putative AdS radius. This will break the scale invariance of the AdS.

The near-brane AdS$_3$ geometries that we investigate are limited to a single intersection, but the solutions  are far from being featureless.  Once the back-reaction is incorporated via a supergravity solution, we find  spikes created by M2 branes ending on  M5 branes that have a profile \cite{Callan:1997kz, Constable:1999ac,Lunin:2007mj}:
\begin{equation}
z ~\sim~    \frac{c}{u^2}   \,.
 \label{profile}
\end{equation}
The right-hand side is generically a harmonic function sourced by the M2's ending on the M5's,   and is thus proportional to $u^{-2}$ given the spherical symmetry.  The steepness of the spike, $c$, is determined by the ratio of the number, $Q_{M2}$, of M2 branes that are pulling and the number, $Q_{M5}$, of M5 branes being deformed. Indeed,  $c \sim \frac {Q_{M2}}{Q_{M5}}$.  However, as we will show, the AdS$_3$ can accommodate any number of  spikes with different steepnesses and the self-similarity of (\ref{profile}) is  what leads to the scale invariance.  Thus we can partition the stack of M5 branes into groups, and choose the number of M2's that end on each group.  This results in different spikes whose steepness is controlled by the value of $\frac {Q_{M2}}{Q_{M5}}$ in each group.

\begin{figure}[h]
    \centering
    \includegraphics[width=.3\textwidth]{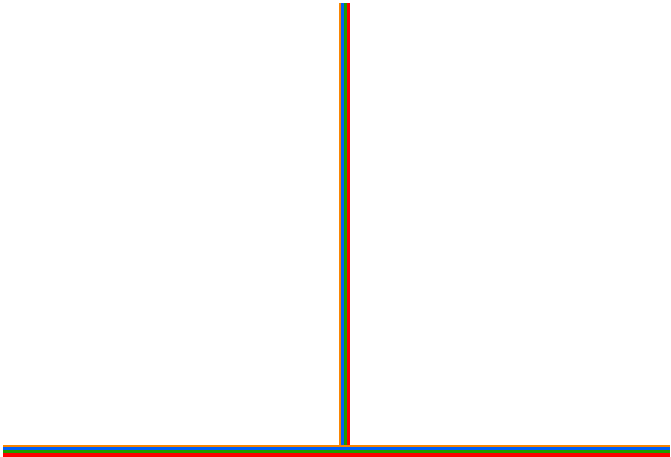} \quad  \raisebox{1.2cm}{\includegraphics[ width=.1\textwidth]{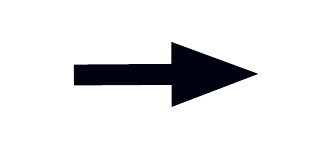} }\quad   \includegraphics[width=.39\textwidth]{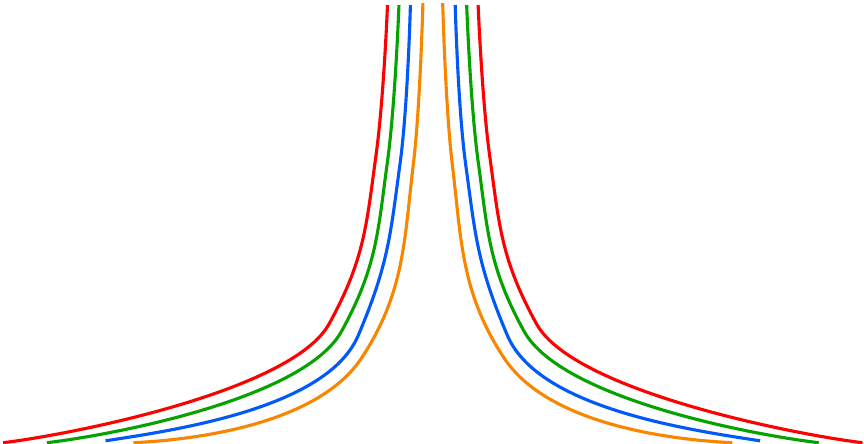}
    \caption{Brane fractionation at a single intersection as a result of back-reacted brane bending.  If the bent branes are self-similar,  preserving a scale invariance, then the near-brane description can result in an AdS geometry.}
    \label{Fractionated2}
\end{figure}

Upon turning on the back-reaction, the different groups spatially separate according to their steepness.  This is depicted in Fig.~\ref{Fractionated2}.  The residues of the function, $G$, control the number of M5's in each group, while the location of the poles of $G$ on the Riemann surface is controlled by the steepness, and thus encodes the M2 charge in each group.    In this way, a ``single intersection'' of non-back-reacted branes actually resolves into a complex of multiple, self-similar intersections whose components localize at different points on the Riemann surface.   At infinity, the brane configuration converges to a stack of coincident M5's in one direction and M2's in another, but as one zooms into the intersection, the M2's and M5's resolve into physically separated  groups.

It is interesting to note that this brane picture describes  the solutions of  \cite{Bachas:2013vza} that have $AdS_4/\mathbb{Z}_2$ asymptotics, as well as the solutions that have $AdS_7$ asymptotics. The $AdS_4/\mathbb{Z}_2$ asymptotics is not surprising: after all, as one goes up the semi-infinite M2 branes, far away from the M5 branes, one expects the geometry they source to be $AdS_4 \times S^7$. However, for the asymptotically-$AdS_7$ geometries one might naively hope  that they describe M2 branes sandwiched between M5 branes \cite{Bachas:2013vza}.  However, as we will discuss in Section \ref{sec:Example},  these solutions are simply a zoom-in limit at the bottom of the spike depicted in Fig.~\ref{Fractionated2}, and hence are degenerate limit of an M2-M5 infinite spike.

In Section \ref{sec:M2M5int} we review the known supergravity solutions for M2-M5 intersections, focussing on the geometry that emerges in the near-intersection limit.  We follow this, in  Section \ref{sec:Example}, with  our primary example because it has all the essential features of a single intersection.   In Section \ref{sec:M2density} we return to the general near-intersection solutions and compute, in detail, the M2-brane charge density function, and evaluate it at all the sources in the example.  This gives us the picture of the M2-M5 mohawk: a collection of self-similar spikes separated according to steepness.   Section \ref{sec:Conclusions} contains some final comments. In Appendix \ref{app:probebranes} we show how M5-brane probes  reproduce a key characteristic of the mohawk.  That is, given a supergravity solution M2-M5 mohawk,  one can analyze the forces experienced by a probe component of the mohawk (an M5-brane with an M2 spike, which in the coordinates of  \cite{Bachas:2013vza} is extended along the $AdS_3\times S_3$ part of the geometry).  We show that the probe has an equilibrium position on the boundary of the Riemann surface depends linearly on the amount of M2 flux along the probe world-volume. This implies that the relation between the M2 charge and the steepness of the spike is that same as that in the spikes whose backreaction gives rise to the background $AdS_3\times S_3\times S_3$ solution. 

\section{General M2-M5 intersections and AdS$_3$ limits}
\label{sec:M2M5int}

\subsection{General M2-M5 intersections}
\label{ss:Lunin}

As described in the Introduction, the brane configuration becomes much clearer in the original formulation of \cite{Lunin:2007mj,Lunin:2008tf}, and we follow the discussion in \cite{Bena:2023rzm,Bena:2024qed}.
The M2 directions correspond to the coordinates  $(x^0, x^1,x^2) = (t,y,z)$, while the coordinates along the  M5 branes are $(t, y,x^3, \dots,x^6)$. The transverse directions are thus $(x^7, \dots,x^{10})$, and will be denoted by  $\vec v \in {\IR}^4$.  Similarly, we use a vector, $\vec u \in \IR^4$, to denote the directions, $(x^3, \dots,x^6)$, transverse to the M2 inside  in M5.    Since we  are going to focus on a single brane intersection, we will impose spherical symmetry in both these $\IR^4$'s.  

The \nBPS{4}  solution then has a metric of the form:
\begin{equation}
\begin{aligned}
ds_{11}^2 ~=~ e^{2  A_0}\, \Big[&  - dt^2 ~+~ dy^2  ~+~  (-\partial_z w ) \, \big( dz ~+~(\partial_z w )^{-1}\,  (\partial_u w)  \, d u \big)^2   \\
& ~+~ e^{-3  A_0} \, (-\partial_z w )^{-\frac{1}{2}}\, \big( du^2  ~+~ u^2 \, d\Omega_3^2 \big)   ~+~ e^{-3  A_0} \, (-\partial_z w )^{\frac{1}{2}}\, \big( dv^2  ~+~ v^2 \, d{\Omega'}_3^2 \big)  
  \Big]\,,
\end{aligned}
 \label{11metric-symm}
\end{equation}
where $u = |\vec u|$, $v = |\vec v|$ and $d\Omega_3^2$, $d{\Omega'}_3^2$ are the metrics of unit three-spheres in each $\IR^4$ factor.  

Based on this, we adopt the frames:
\begin{equation}
\begin{aligned}
e^0 ~=~  &  e^{A_0}\, dt \,, \qquad e^1~=~  e^{A_0}\,  dy \,,  \qquad e^2 ~=~  e^{A_0}\,   (-\partial_z w )^{\frac{1}{2}} \,  \big( dz ~+~(\partial_z w )^{-1}\,   (\partial_u w )  \, d u \big)  \,, \\  e^3 ~=~ &   e^{- \frac{1}{2} A_0} \, (-\partial_z w )^{-\frac{1}{4}}\, du \,,\qquad e^4~=~  e^{- \frac{1}{2} A_0} \, (-\partial_z w )^{\frac{1}{4}}\,  \, dv  \,,\\
e^{i+4}~=~& e^{- \frac{1}{2} A_0} \, (-\partial_z w )^{-\frac{1}{4}}\,    \sigma_i  \,, \qquad e^{i+7}  ~=~ e^{- \frac{1}{2} A_0} \, (-\partial_z w )^{\frac{1}{4}}\, \tilde \sigma_i \,,   \qquad {i = 1,2,3}   \,.
\end{aligned}
 \label{11frames}
\end{equation}
where $\sigma_i$ and $\tilde \sigma_i$ are left-invariant one-forms on the unit three-spheres.

The fluxes are then given by: 
\begin{equation}
C^{(3)} ~=~   - e^0 \wedge e^1 \wedge e^2 ~+~  (\partial_z w )^{-1} \, \big (u^3 \partial_u w \big)  \, {\rm Vol}({S^3}) ~+~  \, \big(v^3 \partial_v w \big) \,  {\rm Vol}({{S'}^3})    \,,
 \label{C3symm}
\end{equation}
where ${\rm Vol}({S^3})$ and ${\rm Vol}({{S'}^3})$ are the volume forms of the unit three-spheres.  

The complete solution is determined by a pre-potential, $G_0(\vec u, \vec v, z)$, that satisfies a generalization of the Monge-Amp\`ere equation:
\begin{equation}
 {\cal L}_{v} G_0 ~=~  ({\cal L}_{u} G_0)\,(\partial_z^2G_0) ~-~ ( \nabla_{\vec u} \partial_z G_0)\cdot  (  \nabla_{\vec u} \partial_z G_0)\,, 
 \label{maze-eqn}
\end{equation}
in which $ {\cal L}_{u} $ and $ {\cal L}_{v}$ denote the Laplacians on each $\IR^4$:
\begin{equation}
{\cal L}_{u} ~\equiv~  \nabla_{\vec u} \cdot \nabla_{\vec u} \,, \qquad {\cal L}_{v} ~\equiv~  \nabla_{\vec v} \cdot \nabla_{\vec v} \,.
 \label{Laps}
\end{equation}
We will refer to (\ref{maze-eqn}) as the maze equation.

Given a solution of the maze equation, the metric and flux functions are given by:
\begin{equation}
w ~=~ \partial_z G_0  \,, \qquad  e^{-3  A_0} \, (-\partial_z w )^{\frac{1}{2}} ~=~ {\cal L}_{v}  G_0 \,.
 \label{solfns}
\end{equation}
Substituting these into  (\ref{maze-eqn}) gives the relation:
\begin{equation}
 e^{-3  A_0} \, (\partial_z w )^{-\frac{1}{2}} ~-~ (\partial_z w )^{-1} \, (\nabla_{\vec u} \,w )\cdot  (\nabla_{\vec u} \, w ) ~=~  - {\cal L}_{u}  G_0  \,.
 \label{reln1}
\end{equation}
%

\subsection{The near-intersection limit}
\label{ss:AdSlim}

The ``near-brane'' limit of M2-M5 intersections is also an implicit part of the work on Janus solutions in M-theory 
\cite{Lunin:2007ab,DHoker:2008lup,DHoker:2008rje,DHoker:2008wvd,DHoker:2009lky, DHoker:2009wlx,Bobev:2013yra,Bachas:2013vza,Bena:2023rzm}. We therefore summarize the key results of \cite{Bachas:2013vza}.

The metric and fluxes have the form: 
\begin{equation}
\begin{aligned}
ds_{11}^2 &~=~   e^{2A} \, \big( \, \hat f_1^2 \, ds_{AdS_3}^2 ~+~ \hat f_2^2 \, ds_{S^3}^2 ~+~ \hat f_3^2 \, ds_{{S'}^3}^2 ~+~ h_{ij} d \sigma^i  d \sigma^j \,   \big)  \,, \\
C^{(3)}  &~=~  b_1\, \hat e^{012} ~+~ b_2\, \hat e^{345} ~+~ b_3\, \hat e^{678}  \,,
\end{aligned}
 \label{DHokerAnsatz}
\end{equation}
where the metrics $ds_{AdS_3}^2$, $ds_{S^3}^2$  and $ds_{{S'}^3}^2$ are the metrics of unit radius on  AdS$_3$ and  the three-spheres and $\hat e^{012}$, $\hat e^{345}$ and $\hat e^{678}$ are the corresponding volume forms. 

The functions $e^{2A}$, $\hat f_j$, $b_j$, and the two-dimensional metric, $h_{ij}$, are, {\it a priori}, arbitrary functions of $(\sigma^1, \sigma^2)$.    One now imposes the BPS equations and the equations of motion so as to determine all the functions in terms of a complex function, $G$, and a real function $h$.

The two dimensional metric is required to be that of a Riemann surface, $\Sigma$,  with K\"ahler potential, $\log(h)$:
\begin{equation}
h_{ij} d \sigma^i  d \sigma^j ~=~ \frac{\partial_w h \partial_{\bar w}   h }{  h^2}  \, |dw|^2    \,,
 \label{KahlerMet}
\end{equation}
where $w$ is a complex coordinate and  $h$ is harmonic:
\begin{equation}
 \partial_w  \partial_{\bar w} h  ~=~ 0   \,.
 \label{harmonic}
\end{equation}
Again, following \cite{Bena:2023rzm}, we introduce the real and imaginary parts of $w$ via:
\begin{equation}
w ~=~  \xi ~+~ i\, \rho \quad \Rightarrow \quad \partial_w ~=~  \coeff{1}{2}\, \big( \partial_\xi ~-~ i\, \partial_\rho \big) \,, \qquad \partial_{\bar{w}}~=~  \coeff{1}{2} \,\big( \partial_\xi ~+~ i\, \partial_\rho \big)  \,.
 \label{wparts}
\end{equation}
It is also convenient to introduce the harmonic conjugate, $\tilde h$, of $h$, defined by requiring that  $-\tilde h + i h$ is holomorphic:
\begin{equation}
 \partial_{\bar w}  (- \tilde h + i h)  ~=~   0 \,.
 \label{harmconj}
\end{equation}
Since $-\tilde h + i h$ is holomorphic we can use them as {\it local} coordinates on the Riemann surface, or, equivalently we can take 
\begin{equation}
- \tilde h ~+~ i h   ~=~  2 \, w ~=~   2 \, ( \xi ~+~ i\, \rho ) \quad \Leftrightarrow \quad h ~=~ -i (w - \bar w)\,.
 \label{simph}
\end{equation}
This (locally)  fixes the metric of the Riemann surface  to be a multiple of the canonical form  Poincar\'e metric:
\begin{equation}
h_{ij} d \sigma^i  d \sigma^j ~=~ \frac{ d\xi^2 ~+~ d\rho^2}{4 \, \rho^2}  \,.
 \label{PoincareMet}
\end{equation}

The metric functions in  (\ref{DHokerAnsatz})  are determined in terms of a complex function $G(w, \bar w)$: 
\begin{equation}
 \hat f_1^{-2}   ~=~   \gamma^{-1} \, (\gamma +1)^2 \,  ( G    \overline G~-~1)  \,, \qquad  \hat f_2^{-2}    ~=~  W_+ \,, \qquad \hat f_3^{-2}   ~=~  W_-  \,,  \ 
 \label{f-functions}
\end{equation}
and
\begin{equation}
\begin{aligned}
  e^{6A}   ~=~  h^2\, ( G    \overline G~-~1)  \, W_+\, W_-   ~=~ \gamma \, (\gamma +1)^{-2}  \,   h^2\,  \hat f_1^{-2}   \,  \hat f_2^{-2}  \,  \hat f_3^{-2}      \,,
\end{aligned}
 \label{eA-function}
\end{equation}
where: 
\begin{equation}
W_\pm  ~\equiv~  | G ~\pm~ i |^2 ~+~ \gamma^{\pm 1} \, ( G    \overline G~-~1)   \,.
 \label{Wpmdefn}
\end{equation}
The parameter, $\gamma$, lies in the range, $-1 \le \gamma \le 1$, and  is a ``deformation'' parameter that determines the underlying exceptional superalgebra $D(2,1;\gamma)  \oplus D(2,1;\gamma)$ \cite{Bachas:2013vza}.   As we will describe below, and as noted in \cite{Bena:2023rzm}, in connecting these near-brane solutions to the M2-M5 intersections described in Section \ref{ss:Lunin} we will fix $\gamma=1$.

The BPS equations require that $G$  satisfies the equation: 
\begin{equation}
 \partial_w   \, G ~=~  \coeff{1}{2}\, (\, G  +  \overline G\,) \,  \partial_w \log(h) \,. 
 \label{Geqn}
\end{equation}
If one writes $G$ in terms of real and imaginary parts, $G = g_1 + i g_2$, and uses the local coordinates (\ref{simph}), this equation becomes:
\begin{equation}
 \partial_\xi  g_1 ~+~  \partial_\rho    g_2 ~=~ 0 \,,  \qquad  \partial_\xi  g_2 ~-~   \partial_\rho    g_1 ~=~ - \frac{1}{\rho} \, g_1\,. 
 \label{gjeqns1}
\end{equation}
These equations imply that one can define potentials, $\Phi$, and $\tilde \Phi$ via:
\begin{equation}
\partial_\xi \tilde \Phi ~=~ - \frac{2}{\rho} \, g_1 ~=~ - \frac{1}{\rho} \, \partial_\rho \Phi     \,, \qquad  \partial_\rho \tilde \Phi ~=~  - \frac{2}{\rho} \, g_2 ~=~  \frac{1}{\rho} \, \partial_\xi \Phi  \,. 
 \label{Phidefns}
\end{equation}
Consistency with  (\ref{gjeqns1}) means that these potentials   must satisfy second-order differential equations:
\begin{equation}
\Big(\partial_\xi^2  ~+~ \partial_\rho^2 ~-~ \frac{1}{\rho} \,  \partial_\rho \Big)   \, \Phi ~=~  0 \,,
 \label{Phieqn}
\end{equation}
and
\begin{equation}
\partial_\xi^2 \tilde \Phi  ~+~ \frac{1}{\rho} \, \partial_\rho \big( \, \rho\,  \partial_\rho \tilde \Phi  \, \big)   \, ~=~  0 \,. 
 \label{tPhieqn}
\end{equation}

The flux functions, $b_i$, are also determined  by $G$ and its potentials:
\begin{equation}
\begin{aligned}
b _1  &~=~  \frac{1}{2}\, \bigg[\,    \frac{\rho \, g_1 }{(G    \overline G~-~1) } + \Phi \bigg] \,, \qquad 
b _2  ~=~    \frac{ 4\,\rho \, g_1}{W_+ } - (\Phi + 2\,\xi ) \,, \\ 
b _3& ~=~   \frac{4\,\rho \, g_1}{  W_-} - (\Phi - 2\,\xi  ) \,,
\end{aligned}
\label{bfunctions}
\end{equation}
which, as we will discuss extensive below, are well-defined locally and up to the addition of constants.

\subsection{The near-brane limit of M2-M5 intersections}
\label{ss:relation}

To summarize, in order to map the solution in Section \ref{ss:AdSlim} to the near-brane limit of the spherically-symmetric brane-intersection of Section \ref{ss:Lunin} one can take:
\begin{equation}
\gamma = 1 \,, \qquad u  = (\mu  \rho)^{\frac{1}{2}}  \, e^{- \frac{1}{4} \,  \tilde \Phi}  \,,   \qquad  v  =  (\mu  \rho)^{\frac{1}{2}}   \,  e^{+ \frac{1}{4} \,  \tilde \Phi}   \,,   \qquad  z  =   \frac{1}{2\rho \mu} \,   e^{\frac{1}{2} \,  \tilde \Phi }  \,    \big(  \Phi +  2 \, \xi \big) \,.
 \label{variables3}
\end{equation}
\begin{equation}
w ~=~- \frac{1}{2 \rho \mu}  \, e^{- \frac{1}{2} \,  \tilde \Phi}  (\Phi- 2 \xi) \,.
\label{wres}
 \end{equation}

It is interesting to note that (\ref{variables3}) and (\ref{wres}) imply
\begin{equation}
 u^2 \, z  ~=~   \coeff{1}{2}\,   \big(  \Phi + 2 \, \xi \big)  \,, \qquad v^2 \, w  ~=~ - \coeff{1}{2}\,   \big(  \Phi - 2 \, \xi \big) \,,
 \label{zwsimp}
\end{equation}
\begin{equation}
\begin{aligned}
e^0 ~=~  &\frac{\mu \, e^{A}}{2\,\sqrt{G \bar G-1}} \, dt \,, \qquad e^1~=~ \frac{\mu \, e^{A}}{2\,\sqrt{G \bar G-1}}\,dy \,, \\
e^2 ~=~  &  - \frac{  \, e^{A}}{\rho \,  \sqrt{(G \bar G-1)\,W_+ \,W_-  }} \, \bigg( \rho\, g_1\, \frac{d \mu}{\mu} +  (G \bar G-1) \big(g_2 \, d\xi - g_1 \, d\rho \big) \bigg) \,, \\
e^3~=~  & \frac{ e^{A}}{2 \,  \sqrt{W_+}} \, \bigg( \frac{d \mu}{\mu}  + \frac{d \rho}{\rho}  +   \frac{1}{\rho} \big(g_1 \, d\xi  + g_2 \, d\rho \big) \bigg) \,, \\
 e^4~=~  & \frac{ e^{A}}{2 \,  \sqrt{W_-}} \, \bigg( \frac{d \mu}{\mu}  + \frac{d \rho}{\rho}  -   \frac{1}{\rho} \big(g_1 \, d\xi  + g_2 \, d\rho \big) \bigg) \,, \\
e^{i+4}~=~  &  \frac{ e^{A}}{2 \,  \sqrt{W_+}} \,    \sigma_i  \,, \qquad e^{i+7} ~=~    \frac{ e^{A}}{2 \,  \sqrt{W_-}} \, \tilde   \sigma_i   \,,   \qquad {i = 1,2,3}   \,.
\end{aligned}
 \label{11frames2}
\end{equation}
%

\section{Primary example}
\label{sec:Example}

Our primary example is designed to produce the near-brane limit of a stack of M2's ending on a stack of M5's.  We will choose a very simple Riemann surface, the Poincar\'e upper half plane, which corresponds to taking $h$ to have a single zero and a pole $w=\infty$:
 \begin{equation}
h=-i(w- \bar{w})\,.
\label{hchoice}
\end{equation}
The choice of $G$ is more complicated.  The most general solution can involve three species of branes: M5 branes with non-back-reacted world-volume along $(t, y,x^3, \dots,x^6)$, M5 branes, (usually denoted M5') along $(t, y,x^7, \dots,x^{10})$ and M2 branes along $(t, y,z)$.    We wish to exclude the M5' branes but want M5 sources, and this determines the pole structure of $G$.  Moreover to get an AdS$_4$ $\times S^7$ geometry, corresponding to semi-infinite M2 branes,   the function $G$ must contain a ``flip-term'' on the boundary of the Riemann surface \cite{Bachas:2013vza}:
 \begin{equation}
h=-i(w- \bar{w}), \ \ \ \ \  G=- \left(i\frac{w-\flippt}{|w-\flippt|}+\sum_{a=1}^{n+1}\frac{\zeta_a{\rm{Im}} (w)}{(\bar{w}-\xi_a)|w-\xi_a|}\right) \,,
\label{Gchoice}
\end{equation}
where the parameters $\flippt$, $\xi_a$ and $\zeta_a$ are real.  Without loss of generality we will also take 
\begin{equation}\label{hera}
\flippt <\xi_1<\xi_2\cdots<\xi_{n+1}.
\end{equation}
The flip term changes the boundary value of $G$ from $+i$ to $-i$ at $w=\flippt$ and $w=\infty$.  We have included the flip parameter, $\flippt$, so that it is  easy to pass to a no-flip solution (in which $G=-i$ on the entire boundary) by taking $\flippt \to -\infty$.   In this way one can easily see that the no-flip solution is a degenerate limit of the solution with a flip.

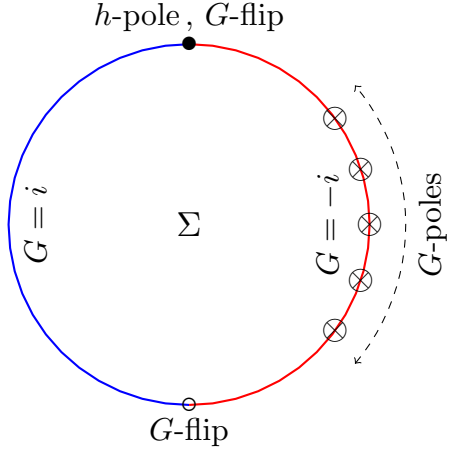
\begin{figure}
\begin{center}
\begin{tikzpicture}[scale=1.2, transform shape]
  \draw [blue,thick,domain=270:90] plot ({2*cos(\x)}, {2*sin(\x)});
    \draw [red,thick,domain=-90:90] plot ({2*cos(\x)}, {2*sin(\x)});
    \draw [black,dashed,<->,domain=-40:40] plot ({2.4*cos(\x)}, {2.4*sin(\x)});
    \node at (2,0) {$\otimes$};
    \node at (1.90211, 0.618034) {$\otimes$};
     \node at (1.90211, -0.618034) {$\otimes$};
     \node at (1.61803, 1.17557) {$\otimes$};
     \node at (1.61803, -1.17557) {$\otimes$};
     \node at (0,2) {$\bullet$};
     \node at (0,-2) {$\circ$};
     \node[rotate=90] at (2.7,0) {\small{$G$-poles}};
      \node at (0,2.3) {\small{$h$-pole\,, $G$-flip}};
       \node at (0,-2.3) {\small{$G$-flip}};
       \node at (0,0) {$\Sigma$};
         \node[rotate=90] at (-1.7,0) {\small{$G=i$}};
          \node[rotate=90] at (1.55,0) {\small{$G=-i$}};
\end{tikzpicture}
\caption{Schematic representation of the Riemann surface $\Sigma$ with the topology of a disc.  The red arc denotes the region on the boundary of $\Sigma$ where $G=-i$ and the blue arc denotes the region on the boundary where $G=i$.   The solid dot $\bullet$ marks the pole in $h$ at $w=\infty$, the poles in $G$ at $w=\xi_a$ are denoted by the symbol $\otimes$ on the red arc and the circle $\circ$ denotes the flip in $G$ at $w=\flippt$. }
  \label{sigmaspace}
  \end{center}
\end{figure}

The poles in  $G$  lie on the boundary ($\rho =0$)  at $w=\xi_a$, and the residue parameters, $\zeta_a$, represent the M5 charges sourced by these poles.  As depicted in Fig.~\ref{sigmaspace}, we have chosen the poles to lie in the interval $\flippt < \xi < \infty$, where $G=-i$.  This implements the choice of only M5 (and not M5') sources.
Metric regularity then requires\footnote{More generally, regularity requires $\zeta_a (\xi_a -\flippt)>0$, but since we only have M5 sources, this means $\zeta_a >0$.}  $\zeta_a >0$.

Using the coordinates (\ref{wparts}),  one can write the real and the imaginary parts of $G=g_1+ig_2$ as:
\begin{equation}
\begin{aligned}
&g_1= \frac{\rho}{\sqrt{(\xi-\flippt)^2+\rho^2}}~-~\sum_{a=1}^{n+1}\frac{\zeta_a\rho(\xi-\xi_a)}{\left((\xi-\xi_a)^2+\rho^2\right)^{3/2}}\,,\\
& g_2= -\left(\frac{(\xi-\flippt)}{\sqrt{(\xi-\flippt)^2+\rho^2}}+\sum_{a=1}^{n+1}\frac{\zeta_a\rho^2}{\left((\xi-\xi_a)^2+\rho^2\right)^{3/2}}\right).
\end{aligned}
\label{gfunctions}
\end{equation} 

One then easily obtains the potentials \eqref{Phidefns}:
\begin{equation}
\begin{aligned}\label{phitildephi}
&\tilde{\Phi}=  2\, \bigg( \log (\flippt\, \rho) ~-~  \log \left( \xi-\flippt+\sqrt{(\xi-\flippt)^2+\rho^2}\right) ~-~ \sum_{a=1}^{n+1}\frac{\zeta_a}{\sqrt{(\xi-\xi_a)^2+\rho^2}}\bigg) \,,\\
&\Phi=2\, \bigg( \flippt~+~ \sqrt{(\xi-\flippt)^2+\rho^2}~+~\sum_{a=1}^{n+1}\frac{\zeta_a(\xi-\xi_a)}{\sqrt{(\xi-\xi_a)^2+\rho^2}}\bigg).
\end{aligned}
\end{equation}
The potentials $\Phi,\tilde{\Phi}$ are defined up to a constant shift.  In the expressions above we have adjusted the constant part of the potentials so that they have a smooth $\flippt\to-\infty$ limit. It is easy to check that in the $\flippt\to-\infty$ limit, one obtains the potentials  discussed in  \cite{Bena:2023rzm}.

To obtain a more geometric picture of the brane layout,  it is useful to express the solution  in terms of the $u,v,z$ coordinates given in \eqref{variables3}:
\begin{align}
u~=~ & \sqrt{\frac{\mu}{|\alpha|}}\left(\xi-\flippt+\sqrt{(\xi-\flippt)^2+\rho^2}\right)^{1/2}e^{ \frac{1}{2} \hat{\Phi}}, \nonumber \\
v~=~&\rho\,\sqrt{\mu |\alpha|}\left(\xi-\flippt+\sqrt{(\xi-\flippt)^2+\rho^2}\right)^{-1/2}e^{- \frac{1}{2} \hat{\Phi}}\,, \label{uvzcoor}\\
z~=~&\frac{|\alpha|}{\mu}\left(\xi-\flippt+\sqrt{(\xi-\flippt)^2+\rho^2}\right)^{-1}e^{- \hat{\Phi}}\Bigg(\xi+\flippt+\sqrt{(\xi-\flippt)^2+\rho^2} +\sum_{a=1}^{n+1}\frac{\zeta_a(\xi-\xi_a)}{\sqrt{(\xi-\xi_a)^2+\rho^2}}\Bigg)\,, \nonumber
\end{align}
where we have defined: 
\begin{equation}
\hat{\Phi} ~\equiv~  \sum_{a=1}^{n+1}\frac{\zeta_a}{\sqrt{(\xi-\xi_a)^2+\rho^2}}\,.
\end{equation}
It is also useful to introduce polar coordinates at the flip point:
\begin{equation}
\xi - \flippt ~=~\lambda^2 \, \cos \theta \,, \qquad \rho  ~=~ \lambda^2 \, \sin \theta \,, 
\end{equation}
which leads to:
\begin{equation}
\begin{aligned}
u~=~ & \sqrt{\frac{ 2\, \mu}{|\alpha|}} \,\, \lambda \, \cos\Big(\frac{\theta}{2}\Big) \, e^{ \frac{1}{2} \hat{\Phi}}\,, \qquad   v~=~ \sqrt{\frac{2\, \mu}{|\alpha|}} \,\, \lambda \, \sin \Big(\frac{\theta}{2}\Big) \, e^{ -\frac{1}{2} \hat{\Phi}},\\
z~=~&\frac{|\alpha|}{2\,\mu}\bigg(\lambda \, \cos\Big(\frac{\theta}{2}\Big) \bigg)^{-2}e^{- \hat{\Phi}}\Bigg{(}2 \,\flippt ~+~ \lambda^2 \, \cos^2\Big(\frac{\theta}{2}\Big) ~+~ \sum_{a=1}^{n+1}\frac{\zeta_a(\xi-\xi_a)}{\sqrt{(\xi-\xi_a)^2+\rho^2}}\Bigg) \,.
\end{aligned}
\label{uvzcoor2}
\end{equation}
Note that $\rho \ge 0$ corresponds to $0 \le \theta \le \pi$.
Also note that 
\begin{equation}
\begin{aligned}
\hat z ~\equiv~ u^2\, z ~=~ & \xi+\flippt+\sqrt{(\xi-\flippt)^2+\rho^2} ~+~ \sum_{a=1}^{n+1}\frac{\zeta_a(\xi-\xi_a)}{\sqrt{(\xi-\xi_a)^2+\rho^2}} \\
&  ~{\to}~ \ \  (\xi+\alpha) ~+~ |\xi-\alpha|~+~ \sum_{a=1}^{n+1} \,  \zeta_a \, {\rm sign}(\xi-\xi_a) \quad {\rm as} \quad \rho \to 0\,.
\end{aligned}
\label{hatz1}
\end{equation}
These coordinate changes reveal much about the configuration.  

First, the brane sources all lie along $\rho =0$, which corresponds to $v=0$.  It is evident from (\ref{11metric-symm}) that $v=0$ defines the origin of the $\IR^4$ transverse to the M2-M5 system.  

The M5 sources are defined by $(\xi = \xi_a, \rho =0)$.  From (\ref{hatz1}), one has at these points:
\begin{equation}
\hat z  |_{\xi=\xi_a ,\rho =0}~=~ 2\xi_a~-~ \sum_{b=1}^{a-1}\zeta_b~+~ \sum_{b=a+1}^{n+1}\zeta_b\,,
\label{hatzval}
\end{equation}
which is a constant on each brane. Indeed, the M5 brane world-volume is defined by $(t,y)$ and the $\IR^4$ with radial coordinate $u$. One  sees from  \eqref{uvzcoor} that along this world volume one has:
\begin{equation}
z ~\sim~  \frac{1}{\mu}\,,  \qquad  u ~\sim~ \sqrt{\mu}\,.
\label{selfsim}
\end{equation}
This shows that the M5 brane is deformed into a ``spike'' in the M2 direction, $z$, with the AdS scale sweeping the radial coordinate in the combined world-volume.  As expected from the results of \cite{Callan:1997kz, Constable:1999ac,Lunin:2007mj}, the spike profile is determined by the harmonic function ($u^{-2}$) sourced by the M2 branes inside the M5 world-volume.  One also sees how the AdS scale invariance arises:  It represents the scaling self-similarity (\ref{selfsim}) of all the spike profiles.  In Appendix \ref{app:probebranes}, we use M5-brane probes to confirm this picture of the solution described by \eqref{Gchoice}.  That is, we show that an M5-brane probe with a world-volume along $AdS_3\times S_3$, and carrying M2 flux, feels no force  when located at a point on the boundary of $\Sigma$ that is determined by the probe's M2 flux. 

As discussed in  the Introduction, in more complicated multi-intersections of branes, the M2 brane profile becomes more complicated, such as in Fig.~\ref{Hjunction}, and the self-similarity is lost.  It is this that defines, and restricts the range of, the near-brane, AdS limit.

The   constant of proportionality in (\ref{selfsim}) determines the steepness of the spike profile, and this is determined by (\ref{hatzval}).  Observe that these values are monotonically increasing in $a$ because $\zeta_a >0$ and  because of (\ref{hera}).  The M5 brane sources are thus a separated collection of spikes (at different values of $\xi_a$) and each such collection is progressively steeper, as depicted in  Fig.~\ref{Fractionated2}.  It is this picture (and Fig.~\ref{mohawk}) that  led to us refer to this configuration as a ``mohawk.''   We will discuss the steepness more extensively in Section \ref{ss:Om1Ex}.

Far from the M5 sources, the function $\hat \Phi$ vanishes, and one has 
\begin{equation}
u^2 ~+~ v^2 ~\approx ~  \frac{ 2\, \mu}{|\alpha|} \,\, \lambda^2  \,.
\label{S7rad}
\end{equation}
This defines the  asymptotic radial coordinate, and  $S^7$, in the $\IR^4 \times \IR^4$ factors of (\ref{11metric-symm}).  This $S^7$ is the Gaussian surface surrounding the M2 branes.

In the limit $\lambda \to 0$, the metric (\ref{DHokerAnsatz}) takes the  form:
\begin{equation}\label{small-lam}
ds^2_{11}~=~ B_1 \, ds^2_{AdS_3}~+~ B_2\,\bigg[d\lambda^2+\lambda^2 \bigg(d \Big(\frac{\theta}{2}\Big)^2+ \cos^2\Big(\frac{\theta}{2}\Big) \, ds^2_{S^3}+\sin^2 \Big(\frac{\theta}{2}\Big)\,ds^2_{S'^3}\bigg)\bigg],
\end{equation}
where
\begin{equation}
B_1~=~\frac{4}{B_2^2}~=~ \left(4\sum_{a=1}^{n+1}\frac{\zeta_a}{(\xi_a-\alpha)^2}\right)^{-2/3}\,.
\end{equation}
Since $0 \le \theta \le \pi$, the factor in square brackets in (\ref{small-lam}) is precisely the metric of flat $\IR^8$.  There are no sources at $\lambda =0$.

As $\lambda \to \infty$, the metric (\ref{DHokerAnsatz}) takes the  form:
\begin{equation}
ds^2~=~ \bigg[ B'_1\, \lambda^4 \, ds^2_{AdS_3} ~+~ B'_2 \, \frac{d\lambda^2}{\lambda^2}\bigg]~+~ B'_2 \bigg[d \Big(\frac{\theta}{2}\Big)^2+ \cos^2\Big(\frac{\theta}{2}\Big) \, ds^2_{S^3}+\sin^2 \Big(\frac{\theta}{2}\Big)\,ds^2_{S'^3}\bigg]\,,
\label{large-lam}
\end{equation}
where 
\begin{equation}
B'_1~=~ \frac{4}{B'^2_2}~=~ \bigg[ 2\,  \bigg(  \sum_{a=1}^{n+1}\,  \zeta_a\bigg)^2  +  4  \sum_{a=1}^{n+1}\, \zeta_a\, (\xi_a - \alpha)\bigg]^{-2/3}\,. 
\label{Bprimes}
\end{equation}
The second factor in (\ref{large-lam}) is exactly the round metric on $S^7$, but now the metric has stabilized to a fixed radius given by (\ref{Bprimes}).  In Section \ref{ss:Om1Ex}, we will show that the term in the square brackets of (\ref{Bprimes}) is a simple multiple of the total M2 charge of the system.

The first factor in (\ref{large-lam})  is a section of the metric on an AdS$_4$.  The easiest way to see this is to note that if $\hat {ds}^2_{AdS_3}$ denotes the metric on a unit global AdS$_3$, then the metric on a unit global AdS$_4$ may be written as:
\begin{equation}
\hat {ds}^2_{AdS_4} ~=~ d\sigma^2 ~+~ \cosh^2 \sigma \, \hat {ds}^2_{AdS_3}\,,
\label{AdS4met}
\end{equation}
where $-\infty<\sigma<\infty$.  In the same way that one can scale global AdS metrics to get Poincar\'e AdS metrics, one can scale (\ref{AdS4met}) to arrive at the first factor of (\ref{large-lam}).  In this sense the latter metric, with $0<\lambda<\infty$, defines an AdS$_4 /\ZZ_2$.

The important point here is that the large-$\lambda$ region of the metric is precisely that of a stack of M2 branes with a radius of curvature determined by the M2-charge.

\section{Computing the M2 Charges}
\label{sec:M2density}

 The equations of motion for $C^{(3)}$ are 
\begin{equation}
d *  F^{(4)} ~=~  - \frac{1}{2}\, F^{(4)}   \wedge F^{(4)}  \,,
 \label{Feqn}
\end{equation}
which imply the existence of an M2-charge density, $C^{(6)}$, defined by:
\begin{equation}
d C^{(6)} ~=~ *F^{(4)}  ~+~    \frac{1}{2}\, C^{(3)}   \wedge F^{(4)} ~+~ {\rm exact} \,.
 \label{C6defn}
\end{equation}
One should note that this is a Page charge:  it is conserved, because of  (\ref{Feqn}),  but it is gauge dependent. 
For the near-intersection limit of Section \ref{ss:AdSlim}, one  can write this as: 
\begin{equation}
d  C^{(6)} ~=~  -d\Omega_1 \, \hat e^{345678} ~+~ d\Omega_2 \, \hat e^{678012}~+~ d\Omega_3 \, \hat e^{345012}\,, 
 \label{C6Omega}
\end{equation}
where the  six-forms, $\hat e^{adcdef}$, are the wedge products of the volume forms introduced in (\ref{DHokerAnsatz}), and the $d \Omega_j$ are one-forms on the Riemann surface, $\Sigma$.     The M2 charge is determined by the first term in  (\ref{C6Omega}), and so we focus on this.

\subsection{The flux functions}
\label{ss:fluxes}

Computing $C^{(6)}$ turns out to involve a few subtleties and so we provide some details here.  
There is a discussion of this in  \cite{Bachas:2013vza}, however we will elucidate this further and make some (minor) corrections.  To facilitate comparison with  \cite{Bachas:2013vza}, we will adopt their notation and conventions  (except we set $\gamma=1$), and use their slightly different normalization of the flux functions. 

We introduce the  potentials, $\hat b_i$: 
\begin{equation}
\begin{aligned}
\hat b _1  &~\equiv~    \frac{ h  \, (G    + \overline G) }{(G    \overline G~-~1) } + 4\, \Phi ~+~ \hat b_1^0 \,, \\
\hat b _2  &~\equiv~  -   \frac{ h  \, (G    + \overline G)}{W_+ } +  (\Phi - \tilde h)~+~\hat b_2^0 \,,  \qquad 
\hat b _3 ~\equiv~   \frac{ h  \, (G    + \overline G)}{  W_-} - (\Phi + \tilde h  )~+~\hat b_3^0 \,,
\end{aligned}
\label{bhatfunctions}
\end{equation}
where we have also included constants of integration\footnote{In \cite{Bachas:2013vza} these constants of integration are denoted $b_j^{0}$. Moreover, the flux functions $b_i$ in \eqref{bfunctions} are related to the ones in \eqref{bhatfunctions} by $b_i=\frac{\nu_i}{c_i^3}\hat b_i$, with $c_1=2, c_2=c_3=-1$, and $\nu_1=\nu_2=-\nu_3=\sigma=1$ \cite{Bena:2023rzm}. 
}, $\hat b_j^{0}$, as they will be important in constructing the M2-charge densities.   

Using  (\ref{C6defn}) and  (\ref{C6Omega}), one finds that the normalized one-form, $d \widehat \Omega_1$, is given by:
\begin{equation}
\partial_w \widehat \Omega_1 ~=~ \frac{ i\, h  \, (G   \overline G-1)^2 }{W_+ W_- }  \, \partial_w \hat b_1 ~-~ \frac{1}{2}\, \big( \hat b_2\, \partial_w \hat b_3 -  \hat b_3 \,\partial_w \hat b_2 \big) ~+~ \partial_w \hat \eta_1 \,, 
 \label{Omega1eqn}
\end{equation}
and its complex conjugate. The function, $\hat \eta_1$, reflects the fact that $d\widehat \Omega_1$ is ambiguous up to an exact piece. With the normalizations of (\ref{bhatfunctions}) and the choices in (\ref{bfunctions}) (see, also, the footnote), it turns out that the original $C^{(6)}$ is obtained from (\ref{C6Omega}) and  $\Omega_1  =  \widehat \Omega_1$.  

\subsection{Non-trivial cycles and smooth fluxes}
\label{ss:smoothflux}

To compute the M2-charges, we need to determine the non-trivial $7$-cycles, and then  choose  $\hat \eta_1$ so that $d \widehat\Omega_1$ is well-defined on each such cycle.  Having done that,  we integrate $dC^{(6)}$ over that $7$-cycle by using Stokes' theorem and  the values $\widehat \Omega_1$ at the endpoints of carefully chosen curves.

The $7$-cycles are either  $S^7$ or $S^4 \times S^3$, and they can be described  using the two $S^3$'s of the geometry and a curve in $\Sigma$ that we will parametrize by $\theta$.  Along this curve, the relevant part of the geometry has the schematic form:
\begin{equation}
ds_7^2     ~=~ d \theta^2 ~+~ k_1(\theta)^2 \, ds_{S^3}^2 ~+~  k_2(\theta)^2 \,  ds_{{S'}^3}^2     \,,
 \label{S7met}
\end{equation}
for some functions $k_j(\theta)$.  One obtains an $S^7$ if $k_1$ vanishes at one end of the $\theta$-curve and $k_2$ vanishes at the other end.  One obtains $S^4 \times S^3$ if one of the $k_i$ remains strictly positive along the curve while the other $k_j$ vanishes at both ends. 

 In the geometry  (\ref{DHokerAnsatz}), the $S^3$'s pinch off at the boundary of $\Sigma$,  ($\rho \to 0$), where $G \to \pm i$.  A curve running between any two points on the boundary of $\Sigma$ thus describes a $7$-cycle, and it is topologically non-trivial if the curve is non-contractible, which happens if the curve  surrounds singular points of $G$ or $h$.  We will only consider situations in which such singular points also lie at the boundary of $\Sigma$.  Because of the symmetry, the integrals over the $S^3$'s are trivial, giving a factor of $4\pi^4$, which we will largely ignore.  The only non-trivial aspect of the calculation is the integral of $d \widehat \Omega_1$ along the curve in $\Sigma$.  If $\widehat \Omega_1$ is continuous along the curve, the integral  reduces to the difference of values of $\widehat \Omega_1$ at the end points of the curve:
\begin{equation}
\int_{X_7} \,   d  C^{(6)}  ~=~ 4\pi^4 \  \widehat \Omega_1\, \Big |_{\xi  = \xi_-}^{\xi  = \xi_+}   \,,
 \label{MonChg}
\end{equation}
where the endpoints of the curve are at $\rho=0$ and $\xi = \xi_\pm$.

As one approaches the boundary of $\Sigma$, the potentials, $\hat b_j$, generically remain finite, and so there is a danger that $d\widehat \Omega_1$ will be singular because the right-hand-side of  (\ref{Omega1eqn}) is finite while a sphere metric is pinching off.  One can adjust the constants, $\hat b_2^0$ and $\hat b_3^0$, so that   $\hat b_2$ vanishes at one point and $\hat b_3$ vanishes at some other point.  In this way, one can use the constants to ensure that $d\widehat \Omega_1$ is well-defined on any topological $S^7$.  However, a problem can arise  for cycles that are topologically $S^4 \times S^3$:  smoothness seems to require that the same $\hat b_j$ must vanish at two different points.   

To resolve this problem, one has to use a non-trivial exact part, $d\hat \eta_1$.   Specifically, if $G\to - i \epsilon$  at both ends of the curve, one can obtain a smooth $d\widehat \Omega_1$ by setting:
\begin{equation}
\partial_w \widehat \Omega_1 ~=~ \frac{ i\, h  \, (G   \overline G-1)^2 }{W_+ W_- }  \, \partial_w \hat b_1 ~-~ \frac{1}{2}\, \big( \hat b_2\, \partial_w \hat b_3 -  \hat b_3 \,\partial_w \hat b_2 \big) ~-~  \frac{1}{2}\, \epsilon \, \partial_w \big( \hat b_2\, \hat b_3 \big) \,. 
 \label{Omega1-smoothed} 
\end{equation}

To see how this works, observe that  as $G\to -i$, the metric on $S^3$ remains finite, while ${S'}^3$ pinches off.  This means that $\hat b_2$ is non-singular (for any $\hat b_2^0$) on the cycle, while finite $\hat b_3$ will lead to a singular $d\widehat \Omega_1$ at the pinch-off points.   Taking $\epsilon =1$ in   (\ref{Omega1-smoothed})   converts the source to $\hat b_2\, \partial_w \hat b_3$, with no ``bare'' $\hat b_3$, thus obviating the effects of a finite $\hat b_3$.    Similarly, as $G \to +i$,  the metric on ${S'}^3$ remains finite and ${S}^3$ pinches off, making finite $\hat b_2$ dangerous, but taking  $\epsilon =-1$ cancels the bare $\hat b_2$ in   (\ref{Omega1-smoothed}).

We now see how this works in detail by computing $\widehat \Omega_1$ explicitly.

\subsection{Computing the flux potential, $\hat \Omega_1$}
\label{ss:Om1}

Again, following  \cite{Bachas:2013vza}, the  solution to   (\ref{Omega1-smoothed}) has the form
\begin{equation}
\begin{aligned}
 \widehat \Omega_1 ~=~ & \frac{h}{2\,W_+} \Big[ h  \, (G   \overline G-1)  ~+~  (\Phi + \tilde h)(G  +  \overline G)\Big]~-~ \frac{h}{2\,W_-} \Big[ h  \, (G   \overline G-1)  ~+~  (\Phi - \tilde h)(G  +  \overline G)\Big]  \\ 
 & ~-~ \frac{1}{2}\,\hat b_2^0 \, \bigg[  \frac{ h  \, (G    + \overline G)}{  W_-} - (\Phi + \tilde h  )  \bigg]~-~ \frac{1}{2}\,\hat b_3^0 \, \bigg[ \frac{ h  \, (G    + \overline G)}{W_+ } -  (\Phi - \tilde h) )  \bigg] \\
 &~-~\tilde h \, \Phi ~+~\Lambda~-~ \frac{1}{2}\, \epsilon \, \hat b_2\, \hat b_3\,, 
\end{aligned}
 \label{Omega1sol} 
\end{equation}
where $\Lambda$ satisfies:
\begin{equation}
\partial_w \Lambda  ~=~ i \, h\, \partial_w \Phi ~-~ 2i \, \Phi\, \partial_w  h   \,. 
 \label{Lambda-defn} 
\end{equation}
The integrability condition for the equation for $\Lambda$ follows from the equation (\ref{Phieqn}) for $\Phi$.  Specifically, one can write (\ref{Lambda-defn}) as
\begin{equation}
\partial_\xi \Lambda  ~=~ 2 \rho \, \partial_\rho \Phi  ~-~  4\, \Phi \,, \qquad  \partial_\rho \Lambda  ~=~ - 2 \rho \, \partial_\xi \Phi\,. 
 \label{Lambda-defn2} 
\end{equation}
Eliminating $\Lambda$ from these equations gives  (\ref{Phieqn}), while eliminating $\Phi$ leads to: 
\begin{equation}
\partial_\xi^2 \Lambda  ~+~ \partial_\rho^2 \Lambda ~-~ \frac{3}{\rho}\,  \partial_\rho \Lambda   ~=~ 0\,. 
 \label{Lambda-eqn} 
\end{equation}

Finally, using (\ref{bhatfunctions}), observe that 
\begin{equation}
\begin{aligned}
 - \frac{1}{2}\, \epsilon \, \hat b_2\, \hat b_3  ~=~ &  \frac{1}{2}\, \epsilon \,   \frac{ h^2  \, (G    + \overline G)^2}{W_+W_- }   ~+~  \frac{1}{2}\, \epsilon  \,  \big(\Phi - \tilde h + \hat b_2^0 \big)\big(\Phi + \tilde h -\hat b_3^0\big) \\ &~-~   \epsilon \,   \bigg[ \frac{h}{2\,W_+} \big(\Phi + \tilde h -\hat b_3^0\big)(G  +  \overline G) ~+~ \frac{h}{2\,W_-} \big(\Phi - \tilde h + \hat b_2^0 \big)(G  +  \overline G)   \bigg] 
\end{aligned}
 \label{b2b3} 
\end{equation}
and hence 
\begin{equation}
\begin{aligned}
 \widehat \Omega_1 ~=~ &  \frac{1}{2}\, \epsilon \,   \frac{ h^2  \, (G    + \overline G)^2}{W_+W_- }  ~+~\frac{i \, h^2 \, (G   \overline G-1)(G   - \overline G) }{W_+ W_-}    \\  
 &~-~  \frac{1}{2}\,(1- \epsilon)  \, \big( \hat b_3^0 - (\Phi + \tilde h)\big) \,  \bigg[ \frac{h}{W_+} \,(G  +  \overline G) -  \big(\Phi - \tilde h\big) \bigg] \\
 & ~-~  \frac{1}{2}\,(1+ \epsilon ) \,  \big( \hat b_2^0+ (\Phi - \tilde h)\big) \, \bigg[ \frac{h}{W_-} \,(G  +  \overline G)   -  \big(\Phi + \tilde h\big)\bigg] \\ 
 & ~-~  \frac{1}{2}\, \epsilon  \,  \big(\Phi^2  - \tilde h^2  \big)  ~-~\tilde h \, \Phi ~+~\Lambda ~-~  \frac{1}{2}\, \epsilon  \,   \hat b_2^0 \, \hat b_3^0 \,.  
\end{aligned}
 \label{Omega1sol-full} 
\end{equation}

Consider the limit of $\widehat \Omega_1$ as $\rho \to 0$.  Recall that, in this limit, one has $G \to \mp i$ which means  $W_\pm \sim {\cal O}(\rho^2)$ and  $W_\mp \to 4$.   It follows that the first two terms in  (\ref{Omega1sol-full}) vanish.  If one has M5 brane sources in  the $G \to - i$ region, as we do in the example of Section \ref{sec:Example}, then, as we discussed above, we use the gauge with $\epsilon =+1$ for $\Omega_1$ to be well-defined.  This leaves: 
\begin{equation}
\begin{aligned}
\widehat \Omega_1\big|_{\rho =0} ~=~ &\Big( \Lambda~-~   \tilde h \, \Phi ~+~  \frac{1}{2}\,   \big(\Phi^2 - \tilde h^2)   ~+~  \hat b_2^0  \,  \big(\Phi +  \tilde h) \Big. \\
\Big.  &~-~  \frac{1}{2}\, \hat b_2^0 \, \hat b_3^0 ~-~  \big( \hat b_2^0+ (\Phi - \tilde h)\big) \frac{h}{W_-}\, (G+ \overline G) \Big)\Big|_{\rho =0}     \,.
\end{aligned}
 \label{Omega1bdry1} 
\end{equation}
%

%
%
Conversely, if the  M5 brane source  lies in  the $G \to + i$ region, one must use the gauge with $\epsilon =-1$, and one is left with: 
\begin{equation}
\begin{aligned}
\widehat \Omega_1\big|_{\rho =0} ~=~ &\Big( \Lambda~-~   \tilde h \, \Phi ~-~  \frac{1}{2}\,   \big(\Phi^2 - \tilde h^2)   ~+~  \hat b_3^0  \,  \big(\Phi -  \tilde h) \Big. \\ 
\Big. &~+~  \frac{1}{2}\, \hat b_2^0 \, \hat b_3^0 ~-~ \big( \hat b_3^0 - (\Phi + \tilde h)\big) \, \frac{h}{W_+}(G + \overline G ) \Big)\Big|_{\rho =0}     \,.
\end{aligned}
 \label{Omega1bdry2} 
\end{equation}
%

%
%

Since we  want to focus on the example in Section  \ref{sec:Example}, we will use (\ref{Omega1bdry1}) and we will drop the constant term  $\hat b_2^0 \, \hat b_3^0$ as this can be absorbed into the definition of $\Lambda$. 

In our example, all the M5 brane sources lie in the $G \to - i$ region and we can actually choose a gauge in which $\widehat \Omega_1$ is globally well-defined. As discussed above, we  need to arrange for $\hat b _2$ to vanish at the boundary where  $G \to +i $.  From (\ref{bhatfunctions}), one therefore must choose:
\begin{equation}
\hat b_2^0  ~=~  - \big(\Phi  - \tilde h  \big)\big|_{\rho =0, \, G \to +i } \,.
 \label{hatb2gauge} 
\end{equation}
The left-hand side of this equation is a constant, while the right-hand side is potentially a function of $\xi$, however we will see that, the right-hand side  is a constant in the $G \to +i $ region ($\xi < \flippt$). Moreover, for $\xi > \flippt$, $W_- \to 4$, and hence the non-trivial term of the second line of (\ref{Omega1bdry1}) vanishes for all $\xi$ to give:
\begin{equation}
\widehat \Omega_1\big|_{\rho =0} ~=~ \Big( \Lambda~-~   \tilde h \, \Phi ~+~  \frac{1}{2}\,   \big(\Phi^2 - \tilde h^2)   ~+~  \hat b_2^0  \,  \big(\Phi +  \tilde h) \Big) \Big|_{\rho =0}\,.
 \label{Omega1bdry3} 
\end{equation}
The value of $\hat b_2^0$-term reflects another gauge choice:  observe that  if one makes a shift $\Phi \to \Phi  + \beta$, where $\beta$ is a constant, then  (\ref{Lambda-defn2}) implies that $\Lambda \to \Lambda  - 4\, \beta\xi  = \Lambda  + 2\, \beta \tilde h$ and therefore
\begin{equation}
\widehat \Omega_1\big|_{\rho =0} ~\to~ \widehat \Omega_1\big|_{\rho =0}  ~+~  \beta\,  \big( \Phi  + \tilde h \big) ~+~  \frac{1}{2}\, \beta^2 ~+~  \hat b_2^0  \,  \beta\,.
 \label{Omega1shift} 
\end{equation}
Thus shifting $\Phi$ by a constant results in a shift of $\hat b_2^0$, and an irrelevant constant shift in $\Lambda$.

\subsection{Computing the flux potential, $\hat \Omega_1$ for the example}
\label{ss:Om1Ex}

For the solution described in Section \ref{sec:Example}, with $\Phi$ given by (\ref{phitildephi}), we find:
\begin{equation}
\begin{aligned}
\Lambda=-4\, \bigg[& \, 2\,\flippt\, \xi + (\xi -\flippt) \sqrt{(\xi-\flippt)^2+\rho^2} ~-~ \alpha^2 \\ 
&  ~-~ \sum_{a=1}^{n+1}\zeta_a \, \bigg( \frac{ \rho^2 }{\sqrt{(\xi-\xi_a)^2+\rho^2}} -2 \sqrt{(\xi-\xi_a)^2+\rho^2}\,\bigg)\bigg]\,, 
\end{aligned}
\label{Lamdaex}
\end{equation}
where we have added a constant term, $4 \alpha^2$, so as to make the  $\flippt \rightarrow - \infty$ limit finite.
Taking $\rho \to 0$ in this example, we find:
\begin{equation}
\begin{aligned}
g_2 \big|_{\rho =0} ~&=~  - {\rm sign}(\xi-\flippt)  \,, \\
\Phi \big|_{\rho =0} ~&=~ 2\,\Big(\flippt + |\xi-\flippt|+\sum_{a=1}^{n+1} \, \zeta_a\, {\rm sign}(\xi-\xi_a)\Big) \,, \\
\Lambda \big|_{\rho =0} ~&=~  -4\, \bigg[ \, 2\,\flippt\, \xi +\, |\xi- \flippt|\,  (\xi -\flippt)  ~-~ \alpha^2~+~ 2\, \sum_{a=1}^{n+1}\zeta_a \, |\xi-\xi_a|\, \bigg]\,.
\end{aligned}
\label{PhiLam-bdry}
\end{equation}
From  (\ref{gfunctions})  one has, as $\rho \to 0$, 
\begin{equation}
G  ~\to~ - {\rm sign}(\xi - \flippt) \, i  .
 \label{Gbdry} 
\end{equation}

We have computed  (\ref{Omega1bdry3})  with the gauge choice  (\ref{hatb2gauge}), which reduces to:  
\begin{equation}
\hat b_2^0  ~=~  - \big(\Phi  - \tilde h  \big)\big|_{\rho =0, \, \xi < \flippt} ~=~  - 2\,\bigg(2\, \flippt   ~-~ \sum_{a=1}^{n+1} \, \zeta_a \bigg)   \,,
 \label{hatb2gauge2} 
\end{equation}
and we find:
\begin{equation}
\begin{aligned}
\widehat \Omega_1\big|_{\rho =0} ~=~ &  2\,  \bigg(  \sum_{a=1}^{n+1} \, {\rm sign}(\xi-\xi_a) \, \zeta_a\bigg)^2  ~+~  8 \,  \sum_{a=1}^{n+1}\, {\rm sign}(\xi-\xi_a)  \, \zeta_a\, \xi_a ~+~ 2\, \hat b^0  \, \sum_{a=1}^{n+1} \, \zeta_a\, {\rm sign}(\xi-\xi_a)   \\  
 & ~-~4\, \big(1- {\rm sign}(\xi-\flippt)\big) \, (\xi -\flippt) \,  \bigg(\frac{1}{2} \,\hat b^0 ~+~ 2\, \alpha ~+~ \sum_{a=1}^{n+1} \, \zeta_a\,  {\rm sign}(\xi-\xi_a)  \bigg)    \,.
\end{aligned}
 \label{Omega1bdryex1} 
\end{equation}
Observe that the second line manifestly vanishes for $\flippt < \xi < \infty$.  Moreover, for $- \infty  < \xi < \flippt$, one has ${\rm sign}(\xi-\xi_a) = -1$, for all $a$,  because of (\ref{hera}), and so the second line vanishes as a result of the gauge choice (\ref{hatb2gauge2}).  Therefore, with our gauge choices,  the result may be written:
\begin{equation}
\widehat \Omega_1\big|_{\rho =0} ~=~  2\,  \bigg(  \sum_{a=1}^{n+1} \, \big( 1 + {\rm sign}(\xi-\xi_a) \big)\, \zeta_a\bigg)^2  ~+~  8 \,  \sum_{a=1}^{n+1}\,\big( 1 + {\rm sign}(\xi-\xi_a) \big)\,   \, \zeta_a\, (\xi_a - \flippt)\,,  
 \label{Omega1bdryex2} 
\end{equation}
where we have adjusted  the  constant term  to recast the expression in a simple form that vanishes as $\xi \to -\infty$, and in which every term is positive (recall that $\zeta_a >0$).

This expression for $\widehat \Omega_1\big|_{\rho =0}$ is globally defined for $-\infty < \xi < \infty$, $\xi \ne \xi_a$,  and it is   {\it locally constant}, as   required by conservation of the Page charge.

Using this, one can compute the M2 charge\footnote{The sign of this charge depends on contour orientation and also does not take into account the negative sign in (4.9) of \cite{Bachas:2013vza}, and so there can be differences in signs that depend upon these conventions. We have chosen to make $Q_{M2, a}$ positive.} sourced at each singular point, $\xi_a$: 
\begin{equation}
Q_{M2, a} ~\equiv~\widehat \Omega_1 \big|_{\rho =0,  \, \xi = \xi_a + \varepsilon }~-~  \widehat \Omega_1 \big|_{\rho =0,  \, \xi = \xi_a - \varepsilon }  
 ~=~  8\, \zeta_a  \,  \bigg(2\, (\xi_a - \alpha) ~+~ \zeta_a  ~+~ 2  \sum_{b=1}^{a-1} \, \zeta_b \bigg) \,,
 \label{Omega1jump} 
\end{equation}
for some small $\varepsilon >0$.  Note that, with our gauge choices, all these charges are positive.

The total M2 charge is given by:
\begin{equation}
Q_{M2 , total} ~=~ \widehat \Omega_1 \big|_{\rho =0,  \, \xi \to + \infty}~-~ \widehat \Omega_1\big|_{\rho =0, \,  \xi \to - \infty} 
 ~=~  8\,  \bigg(  \sum_{a=1}^{n+1} \, \zeta_a\bigg)^2  ~+~   16 \, \sum_{a=1}^{n+1} \, \zeta_a\, (\xi_a - \alpha)    \,,
 \label{Omega1infdiff} 
\end{equation}
and one can easily check see that this is also given by the sum of the contributions (\ref{Omega1jump}), as required by conservation.

Returning to the metric at infinity, (\ref{large-lam})  and (\ref{Bprimes}), we see that radius of curvature is determined by $Q_{M2 , total}$.

\subsection{The brane-intersection mohawk }
\label{ss:Mohawk}

As we discussed in Section \ref{sec:Example}, it is very useful to define the spike-profile coordinate, $\hat z$:
\begin{equation}
 \hat z ~\equiv~ u^2 \, z  ~=~   \coeff{1}{2}\,   \big(  \Phi + 2 \, \xi \big) ~=~ (\xi +   \flippt) ~+~ \sqrt{(\xi-\flippt)^2   ~+~  \rho^2}+\sum_{a=1}^{n+1}\frac{\zeta_a(\xi-\xi_a)}{\sqrt{(\xi-\xi_a)^2+\rho^2}}\,.
 \label{zhatdefn}
\end{equation}
Note that, for $\rho =0$, $\hat z$ is constant for $\xi <  \flippt$, and is linear in $\xi$ for $\xi >  \flippt$ except for jumps at each M5 source by  $2 \zeta_a$.  The heights of these jumps are essentially the M5 charge of the source. 
Moreover, the spike-profile at each source can be written as:
\begin{equation}
 \lim_{\rho \to 0} \Big( \hat z  +  \coeff{1}{2} \, \hat b_2^0 \Big) \Big|_{\xi = \xi_a}  ~=~   2 \,  (\xi_a - \alpha) ~+~\bigg(\zeta_a  ~+~ 2  \sum_{b=1}^{a-1} \, \zeta_b \bigg) ~=~ \frac{Q_{M2, a}}{8\, \zeta_a}\,.
 \label{zhatval}
\end{equation}
This means that the spike-profile at each source  has the form:
\begin{equation}
 \lim_{\rho \to 0} \Big( \hat z  +  \coeff{1}{2} \, \hat b_2^0 \Big) \Big|_{\xi = \xi_a}   ~=~ \frac{Q_{M2, a}}{2\, Q_{M5, a}}\,.
 \label{zhatspike}
\end{equation}
where we have used the fact that the M5 charge is $4\,\zeta_a$ \cite{Bachas:2013vza}.  This equation has a very simple meaning: the  spike is caused by M2's pulling on the M5's, and the steepness of the spike is determined by the number of M2's pulling on the M5's divided by the number of M5's being pulled.   Note that we have written the offset in $ \hat z$ in terms of  $\hat b_2^0$ to reflect the fact that the offset is part of a gauge choice.

One can also write this formula as:
\begin{equation}
 \lim_{\rho \to 0} \hat z \, \big|_{\xi = \xi_a}   ~\sim~ \frac{\big( Q_{M2, a}~-~ \coeff{1}{2} \, \hat b_2^0 \,Q_{M5, a}\big)}{Q_{M5, a}}\,.
 \label{zhatspike-ginv}
\end{equation}
From the perspective of brane intersections, the coordinates $(u,v,z)$, and hence $\hat z$, are universal, and necessarily gauge invariant.  This means that the right-hand side of (\ref{zhatspike-ginv}) is gauge invariant.  Indeed, observe that the combination in the numerator has the form of the gauge invariant brane charge associated with each spike.

\begin{figure}[h]
    \centering
    \includegraphics[width=.4\textwidth]{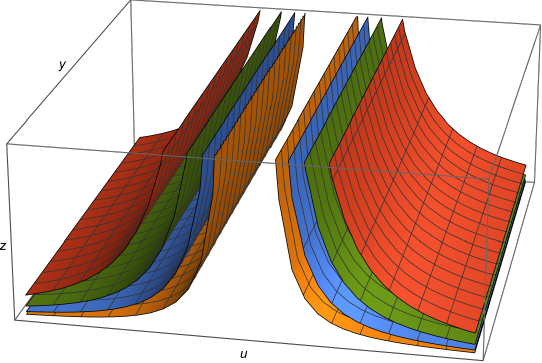}  \qquad \includegraphics[width=.4\textwidth]{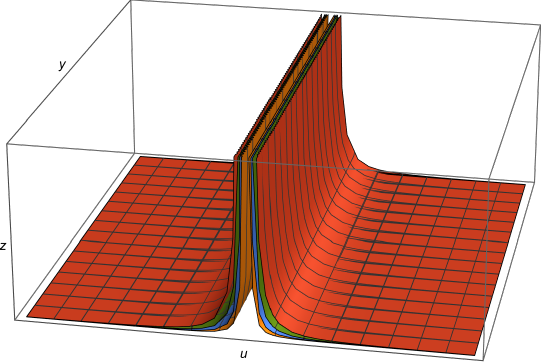}
    \caption{The ``Mohawk:'' the bending of the groups of M5 branes by intersecting M2 branes.  The groups are separated and bent according to the value of  $\frac{Q_{M2}}{Q_{M5}}$.  At infinity, the M2 and M5 branes limit to coincident stacks, as depicted in the figure on the right. The two plots show the mohawk closer in and zoomed out, revealing the separation of the branes and their asymptotic convergence.}
    \label{mohawk}
\end{figure}

As noted in Section \ref{sec:Example}, another very important feature of  (\ref{Omega1jump}) and   (\ref{zhatspike}) is that these quantities increase monotonically with $a$, because $\zeta_a >0$ and  $\xi_{a+1}  > \xi_{a}$.  This leads to an intuitively satisfying picture of the back-reacted brane intersection. Before back-reaction, one has a stack of coincident M2 branes ending on a stack of coincident M5 branes.  One can partition the M5's into groups, with the number in each such group determined by $\zeta_a$.  One is then allowed to choose how many M2's terminate on and dissolve into each of the these groups of M5's.  This is determined by  the number in the numerator of (\ref{zhatspike-ginv}).     The more M2's terminating on each group of M5's, the greater the  bending of the M5 branes: the groups of M5's bend according to the value of each term in (\ref{zhatspike-ginv}).  This causes the groups of M5's to physically separate into distinct localized sources at $\xi = \xi_{a}$, as determined by  (\ref{zhatspike-ginv}).    The sources are ordered according to steepness, with the steepest localized at $\xi = \xi_{n+1}$ and the least steep localized at $\xi = \xi_{1}$.   The M2 charges  thus determine the parameters, $\xi_a$.  This is depicted schematically in     Fig.~\ref{mohawk}.

\section{Final comments}
\label{sec:Conclusions}

Our primary interest in  brane fractionation is to capture the twisted sectors of the CFT's that arise on  coincident stacks of two species of brane.  The standard work-horse for CFT's on brane intersections is the D1-D5 system in which the CFT has a well-understood, weak-coupling limit.  Here we have focussed on M2-M5 intersections because the structure of the solutions on the internal manifold is simpler.  

\begin{figure}[h]
    \centering
  \includegraphics[width=.25 \textwidth]{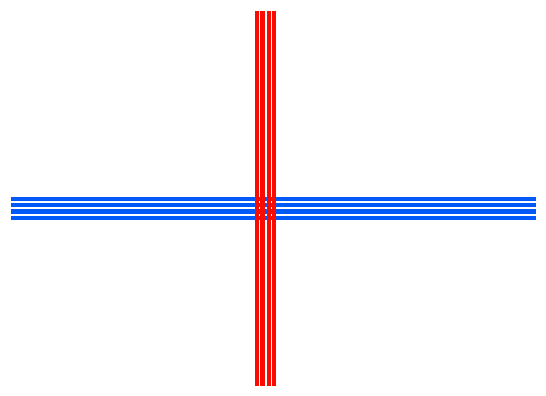}     \raisebox{1.0cm}{\includegraphics[ width=.1\textwidth]{Arrow1.pdf} }      \includegraphics[width=.6 \textwidth]{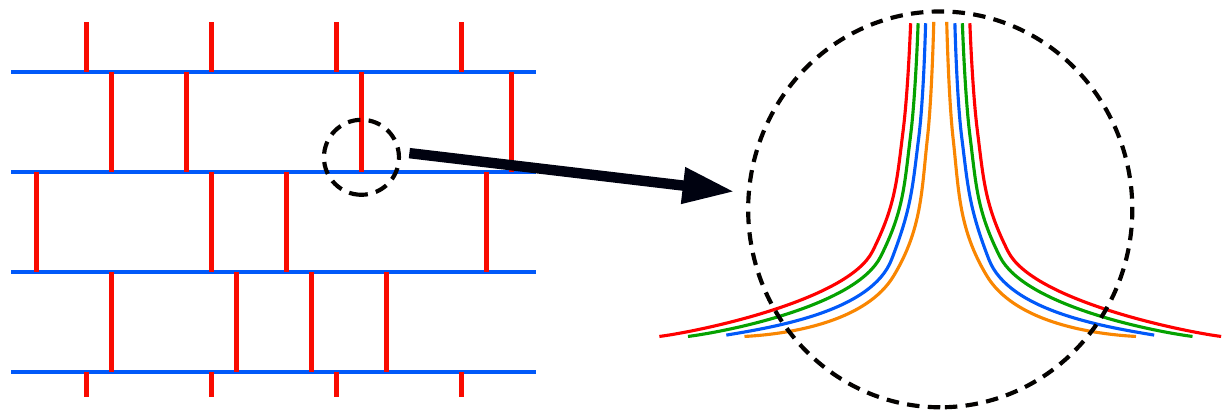}   
    \caption{Brane fractionation.  The first step shows ``standard'' brane fractionation in flat space.  The  second step shows ``second-level''  fractionation in a near-brane limit in which the back-reaction creates self-similar spikes whose scale invariance is  transmogrified into an AdS$_3$ space.  In the first step the  moduli are the positions of each brane segment, while, in the second step, the position of each spike is controlled by the charge ratio in the spike.}
    \label{Fractionated3}
\end{figure}

In the standard picture, the twisted sectors, and  the majority of microstates, emerge from some form of fractionation leading to a ``Higgs Branch.''  In the D1-D5 system one gets $4 N_1 N_5$ scalars  from the instanton moduli space of D1's inside D5's.  For the M2-M5 system, these scalars come from the $4 N_2 N_5$ positions of the fractionated branes depicted in the first part of Fig.~\ref{Fractionated3}.   

To capture this fractionation with supergravity, one has to  fractionate the branes only partially, so that each ``brane segment'' still has a sufficiently large number of branes to produce a significant gravitational back-reaction. We also have to choose the compactification scale to be sufficiently large so that the supergravity approximation is valid. 

In this paper we have shown that brane fractionation can happen at two qualitatively different levels.  The first is the one we just described. However, we have shown that there is a second fractionation, depicted in Fig.~\ref{Fractionated2} and the second part of  Fig.~\ref{Fractionated3}, which occurs at each individual intersection. This fractionation preserves an AdS isometry, creating what we have called the  M2-M5 mohawk. Since this second fractionation occurs within a single AdS$_3$, its holographic  interpretation should be captured by the {\it conformal} field theory dual to  a single brane intersection. 

Consider one such intersection with  $N_2$ M2 branes and $N_5$ M5 branes, and the ``intersection CFT'' that it creates. This can result in many different mohawk configurations that are characterized by all the possible sets of  $\frac{N_{2, a}}{N_{5, a}}$ consistent with the total brane charges. We conjecture that each of these different mohawk configurations corresponds to a ground state of this CFT. It would be interesting to count how many such configurations exist for a total $N_2$ and $N_5$. This is given by the total number of ways one can write families of fractions of the type $\frac{N_{2, a}}{N_{5, a}}$ with $\sum_a N_{2,a}=N_2$ and  $\sum_a N_{5,a}=N_5$. We leave the evaluation of this number and its large-$N$ growth to mathematics aficionados.  More broadly, one would like to obtain a more complete understanding of the underlying CFT and its ground-state structure.

The results presented here suggest a number of very interesting follow-up projects.   First, following on from \cite{Bena:2024qed}, it would be very interesting to add momentum waves to these mohawk solutions so as to obtain \nBPS{8} microstate geometries.  More specifically, the goal of studying supergravity solutions that describe fractionated branes is to see how supergravity can access the twisted sectors of the CFT.  If one can add independent momentum waves to each and every intersection point, then one will obtain a coherent supergravity model of the fractionated black-hole microstructure.  As it was originally envisaged \cite{Bena:2022rna,Bena:2022wpl,Bena:2022fzf,Bena:2023rzm,Bena:2023fjx,Bena:2024qed}, this momentum partitioning was to be done at the ``first level'' of partitioning as depicted in the first part of Fig.~\ref{Fractionated3}.  The challenge in this approach is that the intersecting brane system is governed by a non-linear system of equations.  (Nevertheless, the equations governing the momentum excitations and related fluxes were shown to be linear  in  \cite{Bena:2024qed}.)

The new opportunity presented by this work is the emergence of the second level of fractionation in a near-brane limit,  depicted in the second part of Fig.~\ref{Fractionated3}.  These solutions are simpler,  the background is governed by linear equations, and the brane intersections are characterized via the geometry of a Riemann surface.  These  near-brane geometries and their fractionation will thus provide a simpler setting for the investigation of momentum partitioning at fractionated intersections.


While we believe  our example in Section \ref{sec:Example} is representative, it is, from the perspective of \cite{Bachas:2013vza},  only a small subset of a diverse family of solutions. In particular, there is the parameter, $\gamma$, that determines the  representation of the underlying superalgebra, and there is also the option to consider more general Riemann surfaces with more general flux functions, $G$.  Indeed, there is a discussion of ``Lego pieces'' in \cite{Bachas:2013vza} that suggests that one might be able to plumb together more complicated brane intersections using multiply punctured Riemann surfaces.  

In this paper, we took the Riemann surface to be the entire Poincar\'e half plane.   Moreover, as noted in  \cite{Bena:2023rzm,Bena:2024qed}, the residual supersymmetry  also allows one to include additional M5-brane sources, usually denoted as M5', that  share $(t,y)$ but fill the spatial directions transverse to the original M2 and M5 branes.  These directions are described by $v$  and the sphere $S'{}^3$ in Section \ref{sec:M2M5int}.  We have excluded such M5' sources\footnote{There is, of course, a dielectric distribution of M5' fluxes that, together with the M5 sources, give rise to the M2 charges through the Chern-Simons term.}. (This is why we chose all the sources in Section \ref{sec:Example} to be in the region $G \to - i$.) Based on the analysis in \cite{Bena:2024qed}, we also set the supersymmetry parameter, $\gamma$, to $1$.  

All of this greatly limits the ``Lego pieces'' described in  \cite{Bachas:2013vza}, and excluding M5' brane sources places further limitations.  It would be interesting to see if one can do something more general by freeing up the $\gamma$-parameter, and allowing a more general geometry for the common intersection of the branes.  On the other hand, there has to be a price for taking the near-brane limit and getting an AdS factor in the geometry.  As we have discussed in our example, the scaling of the AdS arises from the self-similarity of the bending of brane intersections. There must be a similar scale invariance in other brane intersections described by the results in  \cite{Bachas:2013vza}, and, as we remarked in the Introduction, this will still  limit  the possibilities. 

The near-brane limits of the intersecting M2-M5 system that can be incorporated  as components of microstate geometries will therefore be a  restricted sub-class  of the families of solutions obtained in   \cite{Bachas:2013vza}.  Nevertheless, we suspect that one can generalize beyond the example presented here, and even in this example we have seen that there is a rich structure to the ``mohawk'' that will prove invaluable to understanding momentum-carrying black-hole microstates.

\vspace{1em}\noindent {\bf Acknowledgements:} 
We would like to thank Costas Bachas and  Eric d'Hoker  for interesting discussions.
The work of IB and NPW was supported in part by the ERC Grant 787320 - QBH Structure. The work of IB was also supported in part by the ERC Grant 772408 - Stringlandscape and by grant NSF PHY-2309135 to the Kavli Institute for Theoretical Physics (KITP).  The work of SC received funding under the Framework Program for Research and ``Horizon 2020'' innovation under the Marie Sk\l{}odowska-Curie grant agreement no 945298. This work was supported in part by the FACCTS Program at the University of Chicago. The work of DT was supported in part by the Onassis Foundation - Scholarship ID: F ZN 078-2/2023-2024 and by an educational grant from the A. G. Leventis Foundation. The work of NPW was also supported in part by the DOE grant DE-SC0011687.
\newpage

\appendix
\vspace{1em}\noindent {\bf  \LARGE Appendices} 

\section{Probe M5-branes}
\label{app:probebranes}
We demonstrated in the main text that the solutions in Section \ref{sec:Example} correspond to a single stack of semi-infinite M2-branes ending on a single stack of M5 branes, which, as one gets closer to their intersection, separate into different M2-M5 spikes depending on the values of $\xi_a$ and $\zeta_a$. The AdS radius, $\mu$, sweeps the radial direction in the combined world-volume. In this Appendix, we would like to underline and elucidate our interpretation of the supergravity geometry by using probes that are  M5 branes with an M2 spike. These probes have nontrivial M5 worldvolume fluxes.  We expect that one can add such a ``spiked'' brane probe to the background determined by \eqref{Gchoice} and,  because of \eqref{Omega1jump}, we anticipate that such probe branes will feel no force when located on the  boundary of the Riemann surfaces at a $\xi$-position that scales linearly with the amount of M2 world-volume charge. We show that both of these expectations are correct. 

Instead of working with the M5-brane action, which is rather complicated \cite{Bandos:1997ui}, we will reduce the M-theory background to a Type-IIA one and evaluate the action of a spiked D4 brane. We start with the metric and fluxes in \eqref{DHokerAnsatz} and use the Poincar\'e $AdS_3$ metric:
\begin{equation}
\begin{aligned}
ds_{11}^2 &~=~ f_1^2\left(\frac{d\mu^2}{\mu^2}+\mu^2\left(-dt^2+dy^2\right)\right) ~+~ f_2^2 \, ds_{S^3}^2 ~+~ f_3^2 \, ds_{{S'}^3}^2 ~+~ f_4^2|dw|^2  \,, \\
C^{(3)}  &~=~  b_1\, \hat e^{012} ~+~ b_2\, \hat e^{345} ~+~ b_3\, \hat e^{678}  \,,
\end{aligned}
 \label{DHokerAnsatz2}
\end{equation}
where we have absorbed the overall $e^{2A}$ factor into the $f_i$'s, which are given by:
\begin{equation}
\begin{aligned}
	f_1^6&=\frac{h^2 W_+W_-}{64(G\overline G-1)^2}\,, \hspace{30pt} f_2^6=\frac{h^2(G\overline G-1)W_-}{W_+^2}\,, \\
	f_3^6&=\frac{h^2(G\overline G-1)W_+}{W_-^2}\,, \hspace{20pt} f_4^6=\frac{|\partial_w h|^6}{h^4}(G\overline G-1)W_+W_- \,.
\end{aligned}
\end{equation} 
In order to go to a Type IIA duality frame, we  reduce the 11-dimensional solution along the $y$ direction. Using the usual  relations between type IIA and 11-dimensional supergravity solutions
\begin{align}
\label{IIA-M}
	ds_{11}^2&=e^{-\frac{2\phi}{3}}ds_{10}^2+e^{\frac{4\phi}{3}}(dx+C_1)^2 \,, \\
	C_3'&=C_3+B_2 \wedge dx \,,
\end{align}
where $x$ is the direction along which we reduce, we arrive at the following type IIA solution:
\begin{equation}
\begin{aligned}
	ds_{10}^2&=-\mu^3f_1^3 dt^2+\frac{f_1^3}{\mu}d\mu^2+\mu f_1 f_2^2 ds_{S^3}^2+\mu f_1 f_3^2 ds_{{S'3}^3}^2+\mu f_1 f_4^2|dw|^2 \,, \\
	C_3&=b_2\, \hat e^{345} ~+~ b_3\, \hat e^{678} \,, \hspace{15pt} B_2=-\mu b_1 dt\wedge d\mu \,, \hspace{15pt} e^{2\phi}=\mu^3 f_1^3\,.
\end{aligned}
\end{equation}

We consider a probe M5-brane extending along $AdS_3\times S^3$ with world-volume M2 flux on it. We will take the worldvolume of the brane to be along the $AdS_3$ factor of the metric, since this is the only way for the probe to preserve the symmetries of the background. However, since we reduce along $y$ we will instead study a D4-brane with world-volume F1 flux along it. We will take the world-volume of the D4 to be parametrized by $(\eta_0,\eta_1, \eta_2,\eta_3, \eta_4)$ with $(\eta_0,\eta_1)$ identified respectively with $t$ and $\mu$ and $(\eta_2,\eta_3,\eta_4)$ identified with the coordinates on $S^3$.
The induced metric on our probe brane is therefore:
\begin{equation}
	d\tilde{s}_5^2=-\mu^3 f_1^3 dt^2+\frac{f_1^3}{\mu}d\mu^2+\mu f_1 f_2^2 ds_{S^3}^2
\end{equation}
and the induced NS-NS and RR fields are 
\begin{equation}
	\tilde{B}_2=-\mu b_1 dt\wedge d\mu \,, \quad \quad \tilde{C}_3=b_2 \hat e^{345}\,. 
\end{equation}
In order to account for the F1 charge, we turn on a world-volume 2-form field of the form:
\begin{equation}
	F_2~=~ \mathcal{F}dt\wedge d\mu~=~ (\partial_t A_\mu - \partial_\mu A_0) \,dt\wedge d\mu\,,
	\label{F2form}
\end{equation}
where the gauge potential, $A$, and the Maxwell field, $\mathcal{F}$, are, in principle, functions of $\mu$ and the Riemann surface coordinates $(\xi, \rho)$. 

It is then straightforward to compute the DBI and WZ actions:
\begin{align}
	S_{DBI}&=-T_4\int\,d^5\eta e^{-\phi}\sqrt{-\det \left( \tilde{G}_{\alpha \beta}+F_{\alpha \beta}+\tilde{B}_{\alpha \beta}\right)} = -T_4 \int dt\, dy \, d\Omega_3 \, f_2^3\sqrt{\mu^2 f_1^6-(\mathcal{F}-\mu b_1)^2} \,, \nonumber \\
	S_{WZ}&=-T_4 \int\, e^{\tilde{B}_2+\tilde{F}_2}\wedge \oplus_n \tilde{C}_n=-T_4 \int  dt 
	\, dy \, d\Omega_3 \,(F-\mu b_1)b_2 -T_4 \int \, \tilde{C}_5\, .
	\label{Action1}
\end{align}
In order to determine $\tilde{C}_5$, we need to use the fact that:
\begin{align}
\label{conventions}
    F_p&=dC_{p-1} \hspace{88pt} \text{for } p<3 \,, \nonumber \\
    F_p&=dC_{p-1}+H_3\wedge C_{p-3} \hspace{25pt} \text{for } p \geq 3\,,  \\
    F_6&=\star F_4 \,, \hspace{20pt} F_8=\star F_2 \,.  \nonumber
\end{align}
Employing these and noting that $C_1=0$ we arrive at 
\begin{equation}
\begin{aligned}
	dC_5=&\star \left( db_2 \wedge \hat e^{345}\right)+\mu b_3 db_1 \wedge dt \wedge d\mu \wedge \hat e^{678} \, \\
	&\star \left( db_3 \wedge \hat e^{678} \right)+\mu b_2 db_1 \wedge dt \wedge d\mu \wedge \hat e^{345} \,,
\end{aligned}
\label{dC5a}
\end{equation}
where $\star$ denotes the 10-dimensional Hodge star operator. The first term of the  expression above gives a $C_5$ along the Riemann surface, $\Sigma$, and $dt\wedge d\mu \wedge \hat e^{678}$, while the second gives a $C_5$ along $\Sigma$ and $dt \wedge d\mu \wedge \hat e^{345}$. Since we only need the pullback of $C_5$ to the D4-brane probe world-volume, we only need the terms of the second line, which give the following contribution:
\begin{equation}
	\mu \frac{f_1^3f_2^3}{f_3^3}\left(\partial_{\xi}b_3\,d\rho-\partial_{\rho}b_3\, d\xi \right)\wedge dt \wedge d\mu \wedge d\Omega_3 + \mu b_2 \left(\partial_{\xi}\,d\xi+\partial_{\rho}\,d\rho \right)\wedge dt \wedge d\mu \wedge d\Omega_3 \,.
\label{dC5b}
\end{equation}
Integrating this expression is quite complicated and since we will eventually only care about the derivative of $C_5$ along the $\xi$ direction we will leave \eqref{dC5b} as it is. 

The conjugate momentum to $\mathcal{F}$ captures the number of F1 strings (or equivalently M2 branes) that form the spike that ends to the D4 (or M5) branes:
\begin{equation}
	\Pi=\frac{\partial \mathcal{L}}{\partial {(\partial_t A_\mu)}}=\left(-b_2+\frac{f_2^3\left(\mathcal{F}-\mu b_1 \right)}{\sqrt{\mu^2 f_1^6-\left(\mathcal{F}-\mu b_1 \right)^2}}\right) \,.
\label{Cmomentum}
\end{equation} 
The Hamiltonian density is now easily obtained:
\begin{equation}
	\mathcal{H}~=~ \Pi  \, (\partial_t A_\mu) ~-~ \mathcal{L}~=~ -\mu b_1 b_2+f_2^3\frac{\mu b_1 \left(\mathcal{F}-\mu b_1 \right)+\mu^2 f_1^6}{\sqrt{\mu^2 f_1^6-\left( \mathcal{F}-\mu b_1 \right)^2}}+\tilde{C}_5\,.
\end{equation}

The equation of motion obtained by varying the action, (\ref{Action1}), with respect  to $A_0$, yields $\partial_{\mu}\Pi=0$. This is satisfied if one chooses $\mathcal{F}=\mu F(\rho, \xi)$. Solving for $F$ in \eqref{Cmomentum}, we obtain
\begin{equation}
	F~=~ b_1 \pm \frac{f_1^3 |b_2+\Pi|^2}{\sqrt{f_2^6+\left(b_2+\Pi\right)^2}} \,,
\label{F}
\end{equation}
which we use to express the Hamiltonian only in terms of $\Pi$:
\begin{equation}
\mathcal{H}/\mu~=~ -b_1 b_2+f_1^3 \sqrt{f_2^6+\left(b_2+\Pi\right)^2}\pm b_1|b_2+\Pi| +\tilde{C}_5 \,.
\label{finalH}
\end{equation}
This Hamiltonian is a function of the Riemann-surface coordinates $(\xi,\rho)$, but we are interested in putting probes on $\partial \Sigma$ and so we will take its $\rho \rightarrow 0$ limit. We will then interpret it as a potential in the $\xi$ direction and find its minima for a given value of $\Pi$ in a given background geometry. The  expression in \eqref{finalH} has two possible forms depending on which solution of $F$ we choose in \eqref{F} and on whether $b_2+\Pi$ is less or greater than zero. This choice depends on whether we add M2 or anti-M2 charge  to the M5-world-volume. For the particular example we will study here we will take the minus solution and assume that $b_2+\Pi<0$ to arrive at
\begin{equation}
	\mathcal{H}/\mu=\Pi \, b_1 + f_1^3 \sqrt{f_2^6+\left(b_2+\Pi\right)^2} +\tilde{C}_5 \,.
\end{equation}
Moreover, we will express $b_i$ in terms of $\hat b_i$ (see the discussion below \eqref{bhatfunctions}) and we will factor out $\frac{\nu_3 \sigma}{c_1^3c_2^3}$ to simplify our computation. Keeping only the $d\xi$ terms in \eqref{dC5b} and noting that $c_1 c_2 c_3 f_1 f_2 f_3=\sigma h$ we arrive at:
\begin{equation}
	\partial_{\xi} \mathcal{H}=\frac{\nu_3 \sigma}{c_1^3c_2^3} \partial_{\xi}\hat{\mathcal{H}}= \frac{1}{8}\left(8\,\partial_{\xi} \left(f_1^3\sqrt{f_2^6+(\Pi-\hat b_2)^2}\right)+(\Pi-\hat b_2)\, \partial_{\xi}\hat b_1 -\frac{h^3}{f_3^6}\partial_{\rho}\hat b_3 \right) \, .
\label{delH} 
\end{equation}

\begin{figure}[h!]
	\begin{subfigure}[h!]{0.5\linewidth}
		\includegraphics[width=\linewidth]{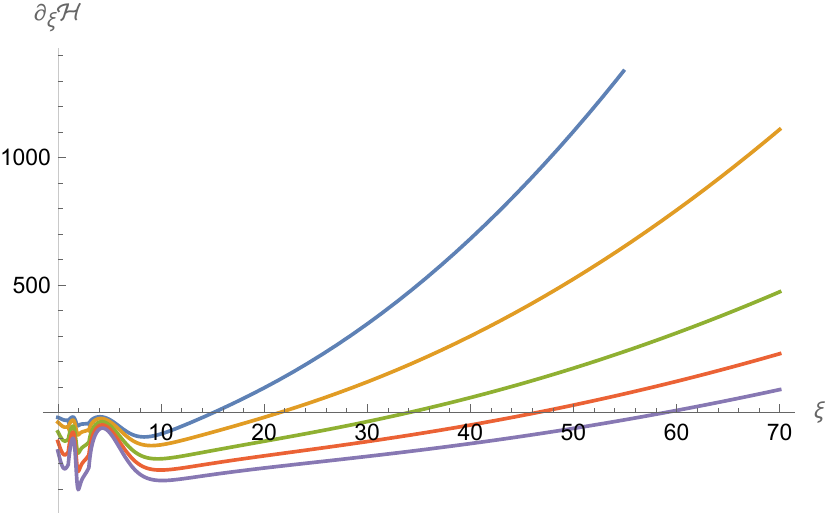}
		\caption{Two-pole solution}
	\end{subfigure}
	\hfill
	\begin{subfigure}[h!]{0.5\linewidth}
		\includegraphics[width=\linewidth]{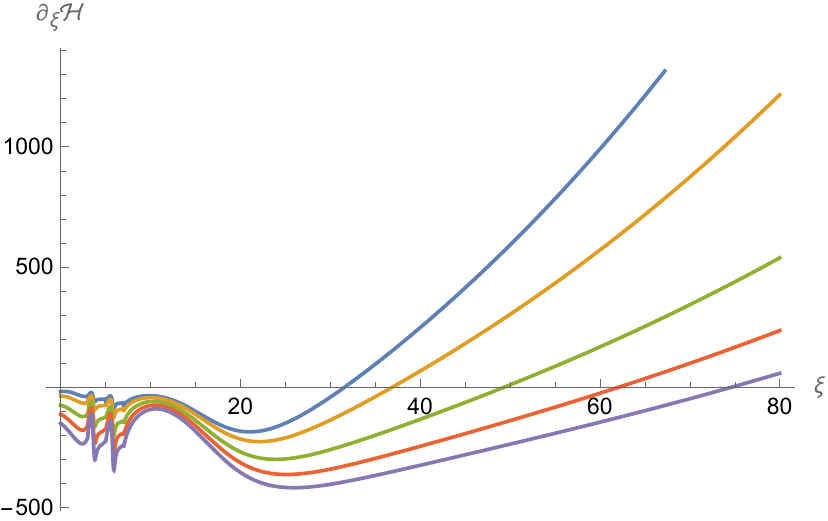}
		\caption{Three-pole solution}
	\end{subfigure}%
	\caption{Solution with two poles in $G$ with parameters $(\xi_1,\xi_2,\zeta_1,\zeta_2,\hat b_0^2)=(1,3,1,1,4)$ (a) and solution with three poles in $G$ with parameters $(\xi_1,\xi_2,\xi_3,\zeta_1,\zeta_2,\zeta_3,\hat b_0^2)=(3,5,7,1,2,3,12)$ (b). In both graphs $\Pi$ takes the values $(-25,-50,-100,-150,-200)$ from left to right. } 
	\label{probeplots}
\end{figure}

In Fig.~\ref{probeplots} we depict the function $\partial_{\xi}\mathcal{H}$ for various values of $\Pi$ and for solutions with two and three poles in $G$. By explicitly computing the zeros of this function we observe that they follow a linear growth in $\Pi$. In particular,  we find that 
\begin{equation}
	\xi_0 \simeq -\frac{1}{4}\Pi+c \,, 
\label{linearPi}
\end{equation}
where $\xi_0$ denotes the minimum of $\mathcal{H}$ and $c$ is a constant that depends on our gauge choices and the parameters of the particular solution we are probing.

The relation, \eqref{linearPi}, is expected since $\Pi$ roughly corresponds to $\sim Q_{M2}/Q_{M5}$ and if we think of our probe as another spike in the solution we are examining, we expect from \eqref{Omega1jump} that its position on the $\xi$ axis will scale with $Q_{M2}$ in the following way\footnote{The signs of the M2 charges here are the opposite of those in the main text, but this  simply reflects our choices  of convention.}: 
\begin{equation}
	\frac{Q_{M2,0}}{Q_{M5,0}}\sim -\frac{16\zeta_0}{4\zeta_0}\xi_0 \quad \Rightarrow \quad \xi_0 \sim -\frac{1}{4}\Pi \,. 
\end{equation}

For the values of $\Pi$ and solution parameters we used in Fig.~\ref{probeplots} there are only minima to the right of the background spikes. However, for smaller values of $\Pi$ one can generally find a minimum also to the left of the spikes, and for sufficient separation between the location of the various poles it is also possible to find a minimum in between them.   We therefore find that the equilibrium positions of our probes match precisely with the supergravity solution and our interpretation of it.


\begin{adjustwidth}{-1mm}{-1mm} 

\bibliographystyle{utphys}      

\bibliography{references}       

\end{adjustwidth}


\end{document}